\documentclass{article}       

\usepackage{amssymb,amsmath}
\usepackage{graphicx}
\usepackage{hyperref}
\usepackage{array}
\usepackage{multirow}
\usepackage{xspace}
\usepackage{booktabs}
\usepackage{siunitx}
\usepackage{todonotes}
\usepackage{subcaption}
\usepackage{caption}
\usepackage[numbers]{natbib}
\usepackage{bold-extra}
\usepackage{enumerate}

\usepackage{longtable,tabulary}

\newcommand{\rqone}{\textbf{RQ1}\xspace}
\newcommand{\rqtwo}{\textbf{RQ2}\xspace}
\newcommand{\rqthree}{\textbf{RQ3}\xspace}
\newcommand{\rqfour}{\textbf{RQ4}\xspace}
\newcommand{\rqfive}{\textbf{RQ5}\xspace}

\newcommand{\psnumber}{65\xspace}
\newcommand{\gen}{\textbf{Ind}\xspace}
\newcommand{\lsp}{\textbf{Dep}\xspace}

\newcommand{\oldl}{\textbf{Exist}\xspace}
\newcommand{\newl}{\textbf{New}\xspace}
\newcommand{\frm}{\textbf{FRM}\xspace}

\newcommand{\semifrm}{\textbf{semiFRM}\xspace}

\newcommand{\model}{\textbf{Mod}\xspace}
\newcommand{\modexe}{\textbf{ModExe}\xspace}

\newcommand{\dm}{\textbf{Dom}\xspace}
\newcommand{\dmi}{\textbf{DomI}\xspace}

\newcommand{\eval}{\textbf{Eval}\xspace}
\newcommand{\neval}{\textbf{NEval}\xspace}
\newcommand{\proc}{\textbf{Proc}\xspace}
\newcommand{\dec}{\textbf{Dec}\xspace}
\newcommand{\act}{\textbf{Act}\xspace}
\newcommand{\art}{\textbf{Art}\xspace}
\newcommand{\hyb}{\textbf{Hyb}\xspace}

\begin{document}

\title{What is a Process Model Composed of? \\ {\large A Systematic Literature Review of Meta-Models in BPM}}


\author{Greta Adamo$^{1,2}$ \quad
        Chiara Ghidini$^1$ \quad
				Chiara Di Francescomarino$^1$\\
$^1$ Fondazione Bruno Kessler, Via Sommarive, 18, 38123 Povo, TN, Italy\\
$^2$ Universit\`a di Genova, Via Opera Pia, 13 16145, GE, Italy
}

\date{}

\maketitle

\begin{abstract}
Business process modelling languages typically enable the representation of business process models by employing (graphical) symbols. These symbols can vary depending upon the verbosity of the language, the modelling paradigm, the focus of the language, and so on. 
To make explicit the different constructs and rules employed by a specific language, as well as bridge the gap across different languages, meta-models have been proposed in literature. These meta-models are a crucial source of knowledge on what state-of-the-art literature considers relevant to describe business processes.
The goal of this work is to provide the first extensive \textit{systematic literature review} (SLR) of business process meta-models. This SLR aims at answering research questions concerning: (i) the kind of meta-models proposed in literature; (ii) the recurring constructs they contain; (iii) their purposes; and (iv) their evaluations.
The SRL was performed manually considering papers automatically retrieved from reference paper repositories as well as proceedings of the main conferences in the Business Process Management research area. 
Sixty-five papers were selected and evaluated against four research questions. The results indicate the existence of a reasonable body of work conducted in this specific area, but not a full maturity. In particular, in answering the research questions several challenges have (re-)emerged for the Business Process Community, concerning: (i) the type of elements that constitute a Business Process and their meaning; (ii) the absence of a (or several) reference meta-model(s) for the community; (iii) the purpose for which meta-models are introduced in literature; and (iv) a framework for the evaluation of the meta-models themselves. Moreover, the classification framework devised to answer the four research questions can provide a reference structure for future descriptive categorisations.

\end{abstract}

\section{Introduction}
\label{introduction}

Business process modelling languages (BPMLs) typically enable the representation of business processes via the creation of process models, 
 which are constructed using the elements and graphical symbols of the BPML itself. A process \emph{model} is a conceptual/abstract representation of a business process, whose goal is to describe or prescribe a \textit{real} process by specifying how the process should/could/might be performed. 
The different constructs and rules employed by a specific BPML to create models are contained in the business process meta-model (BPMM). By quoting \citet[pg.~76]{DBLP:books/daglib/0029914}
\begin{quote}
``Models are expressed in metamodels that are associated with notations, often of graphical nature. For instance the Petri net metamodel consists of places and transitions that form a directed bipartite graph. The traditional Petri net notation associates graphical symbols with metamodel elements. For instance, places are represented by circles, transitions by rectangles, and the graph structure by directed edges.''
\end{quote}

Due to the number of BPMLs available in literature, a number of associated meta-models exist. These meta-models can vary greatly, reflecting the expressive power of the language, its specificities in terms of the specific sub-domain it may focus at, or the modelling paradigm to which the BPML adheres to. 
Meta-models are also defined in literature independently from specific BPMLs with the aim of ``navigating'' across the different BPMLs, bridging the gap across them, fostering a common ground across different notations, and promoting interoperability. This makes the overall number of meta-model proposals increase even more. 
Besides the specific purposes for which they are introduced, meta-models are a crucial source of knowledge on the constructs and rules that state-of-the-art literature considers relevant to model (and thus describe) business processes; yet a detailed analysis of business process meta-models described in literature is still absent. 

Moreover, the growth of approaches and tools aiming at supporting business processes in a multi-perspective manner by looking beyond  the control-flow perspective and including other dimensions, such as the data, organisational and goal oriented ones, shows that the time is now ripe to focus on an investigation of different types of process constructs also at the conceptual level. Indeed, even though all the most popular definitions of business process contain aspects that go beyond the control flow, some of them are still neglected, or not clearly described, in state-of-the-art meta-models.

The goal of this work is to provide a significant extension of the \textit{Systematic Literature Review (SLR)} of business process modelling language meta-models in the BPM (Business Process Management) field contained in \cite{adamo2019whats}. The SLR aims at identifying, categorising, and describing works related to business process meta-models. It focuses on works in the BPM research area and has been driven by four research questions concerning (i) the kind of meta-models proposed in literature; (ii) the recurring constructs they contain; (iii) the purpose(s) of the proposed meta-models; and finally (iv) their evaluations.
In answering the research questions, several challenges have emerged, or re-emerged, for the Business Process Community, concerning: (i) the type of elements that constitute a Business Process and their meaning; (ii) the absence of a (or several) reference meta-model(s) for the community; (iii) the purpose for which meta-models are introduced in literature; and (iv) a framework for the evaluation of the meta-models themselves. Moreover, the classification framework devised to answer the four research questions can provide a reference structure for future descriptive categorisations. The paper extends and encompass the work in~\cite{adamo2019whats} by explicitly including in the search papers about the Petri nets~\cite{Murata89} and DECLARE~\cite{Pesic:2007:DFS:1317532.1318056} modelling languages as well as by adding workflows to the search along with the term process model.          

The paper is organised as follow: in Section~\ref{sec:method} the method employed to perform the SLR is presented, by describing both the planning and the conducting of the review. Special emphasis is given in this section to the definition of the research questions (Section~\ref{sec:RQ}), and to the protocol of review with the description of inclusion and exclusion criteria (Section~\ref{sec:protocol}). The results of our data (papers) collection and selection are given in Section~\ref{sec:results}, and a brief summary of the \psnumber selected papers we retained for answering to the research questions is given in Section~\ref{sec:summaryofpapers}. A detailed answer to the four research questions is provided in Section~\ref{sec:RRQs}, followed by an extensive discussion in Section \ref{sec:discussion}. Final remarks and conclusions are presented in Section~\ref{sec:conclusion}. 

\section{Method}
\label{sec:method}

\begin{figure}[t]
\centering
\includegraphics[width=.80\textwidth]{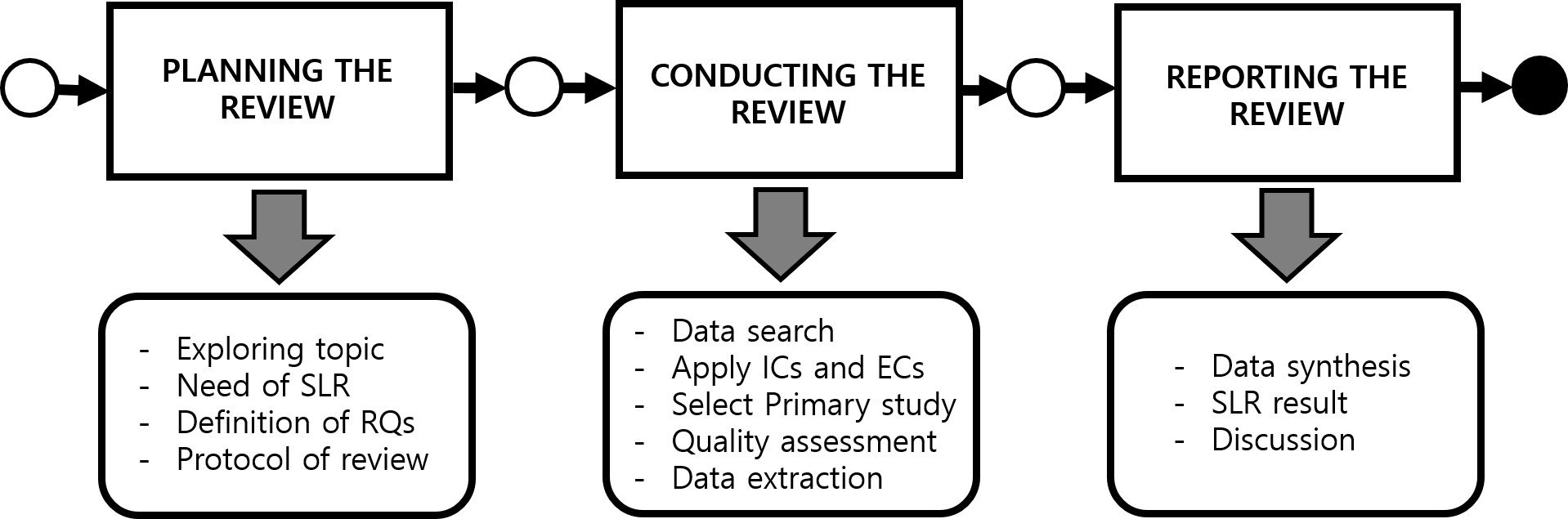}
\caption{Method used for the SLR.}
\label{fig:method}
\end{figure}

The SLR presented in this paper follows the guidelines for conducting a SLR proposed in~\cite{kitchenham2004procedures, Kitchenham07guidelinesfor}. Following these guidelines, a SLR is divided in three pivotal phases, graphically summarised in Figure~\ref{fig:method}: \emph{planning} the review; \emph{conducting} the review; and \emph{reporting}. 

The remaining of this section is devoted to the description of the \emph{planning} of this SLR with particular emphasis to: a brief description of the aspects of the topic relevant to the SLR, the motivations for performing it, the development of the research questions, and the definition of the specific protocol for review adopted. The \emph{conduction} of the review, and its results are described in the next three sections, while this paper, and the additional material linked in the paper, constitutes the SLR \emph{report}. 

\subsection{Background}
\label{sec:background}

The most modern and popular definition of business process is likely the one provided by \citet{DBLP:books/daglib/0029914}:
\begin{quote}
	\label{def:process}
	``a set of activities that are performed in coordination in an organizational and technical environment. These activities jointly realize a business goal. Each business process is enacted by a single organization, but it may interact with business processes performed by other organizations.''
\end{quote}
Thus, business processes are composed of a set of interacting participants (activities, organisational roles, goals just to mention a few\footnote{A comparative analysis of elements belonging to different business process modelling notations can be found in \cite{Adamo:2017:AIIA}}) which are typically captured by business process models specified in a business process modelling language. These languages, and the underlying modelling methodologies, often steam from the conceptual modelling field, whose aim is to identify, analyse, capture, and describe the basic concepts and features of a domain (\textit{universe of discourse}) \cite{32d47c9b496748ff911fb88726992462}. 

Within the last 15 years, an increasing effort has been spent
in providing business process modelling techniques, methodologies, as well as tools and languages for the representation of business process models \cite{DBLP:journals/csur/MiliTJLEE10,AGUILARSAVEN2004129,Ghidini12serafini:modeling}. Focusing on the modelling languages, it is easy to observe their great variability in terms of constructs and rules they offer to compose process models. These differences have different causes ranging from the different expressive powers of the languages, to the specific sub-domain the language may focus at, or even the specific modelling paradigm and approach the modelling language adheres to. Think for instance at the different constructs and rules offered by languages such as BPMN 2.0\footnote{\url{https://www.omg.org/spec/BPMN/2.0/About-BPMN/}}, UML-AD\footnote{\url{https://www.omg.org/spec/UML/About-UML/}}, EPCs~\cite{scheer2013aris}, Petri nets~\cite{Murata89}, DECLARE~\cite{Pesic:2007:DFS:1317532.1318056}, YAWL~\cite{YAWL:2010aa}, and CMMN\footnote{\url{https://www.omg.org/spec/CMMN/About-CMMN/}}, partly due to their declarative vs imperative, activity- vs artefact-centric nature. 
We introduce here some modelling paradigms that will be mentioned throughout the SLR. 

\paragraph{Imperative vs. Declarative Paradigms} 
Imperative BPMLs enable designing process models by specifying the allowed flows of activities.
 Thus, imperative languages such as BPMN, EPC, UML-AD, YAWL, and Petri nets provide particular elements to denote the start and the end of a process \cite{Adamo:2017:AIIA} and  force the production of process models that specify all the possible ways the control flow moves from the start towards the end element. It has been shown that this kind of paradigm is suitable for predictable processes with few variations, but that it is not so effective in situations in which there are many variabilities \cite{DBLP:conf/caise/GiacomoDMM15}.

Declarative BPMLs, such as DECLARE and CMMN, have hence been proposed. As reported in~\cite{DBLP:conf/caise/GiacomoDMM15}, these languages allow modellers to (only) specify constraints on the allowed flows, that is, unless a flow does not satisfy the provided constraints, it is allowed. 
 As a consequence, declarative languages, such as DECLARE, focus on how to express relationships (constraints) between specific process participants rather than modelling a comprehensive view of the control flow with e.g., a well defined start and a well defined end\footnote{As an example DECLARE provides constructs for a first/last  activity, but it does not force neither suggest a process model should always include
 them.} \cite{DBLP:conf/caise/GiacomoDMM15}. 
Traditional BPMLs follow the imperative paradigm.

\paragraph{Activity-centric vs. Artefact-centric Paradigms} 
Activity-centric BPMLs see the process control flow as a series of activities that enable the process to move from the start towards the end construct.
 Thus, as reported in~\cite{DBLP:conf/bpm/0001W13}, languages such as BPMN, UML-AD, YAWL, Petri nets, and DECLARE use activities and control structures (i.e., gateways) as
primary modelling elements, while considering data objects as secondary
 components, often used as pre- and post-conditions for the execution of an activity, or as decision indicators in case of control structure conditions~\cite{DBLP:conf/bpm/0001W13}.

 Differently, artefact-centric process modelling, as for instance CMMN, considers data objects and their life-cycles as primary
 modelling elements, and activities are of importance as they participate to an object change of state. This, more recent, paradigm has been developed and proved useful in scenarios where the flow of the process is originated from the data objects, as for example in case of 
 manufacturing processes \cite{10.1007/978-3-540-76848-7_10}. Most traditional BPMLs follow an activity-centric paradigm.

\subsection{Need for the systematic literature review}
\label{sec:needforSLR}

Modern definitions of business processes, such as the one provided in the previous section, show that the time is now ripe to  investigate, at the conceptual level, process constructs that encompass the control-flow perspective and include other dimensions, such as the data, organisational and goal oriented ones. This is also emphasised by the growth of different modelling languages and approaches aiming at supporting business processes in a multi-perspective manner. A commonly agreed broad view on business processes, with clear and shared definitions of business process elements such as resources, data needed and produced by activities, different types of events, and so on, would be crucial to foster the communication and the data compatibility among information system procedures and data structures designed and described using different modelling paradigms and notations. Nonetheless, such commonly agreed broad view is still not present in the BPM literature. Instead a number of different meta-models are present in literature. These meta-models vary greatly ranging from very general ones, to meta-models tailored to a specific business process modelling language. Certain differences may be explained in terms of
the different perspectives on Business Processes expressed by different meta-models. Nonetheless, conflicts and misalignments on core elements, such as e.g., the definition of what a resource is, appear to be more problematic for the community.\footnote{While we do not necessarily advocate a quest for ``the'' single Business Process meta-model, we nonetheless believe that different views on a business process should be represented by reference meta-models and, more important, the relations between these meta-models should be clear and well understood.} A way to plan the quest for such a unifying broad view may follow a top-down approach. This would lead to the definition of yet another meta-model and to an attempt to impose this meta-model as a reference one. A different way, which we somehow advocate in this paper, is to consider the rich knowledge on the constructs and rules that state-of-the-art literature considers relevant to describe business processes contained in already available \emph{meta-models} of BPMLs. 
Indeed, \emph{meta-models} are used to capture the types of elements included in a notation and the way these elements can be related to each other. They can also make explicit the level of granularity of a business process (e.g., instance level, model level), or the specific sub-domain (dimension) they focus on (e.g., organisation-oriented, information-oriented, and behaviour-oriented). 

To be able to use this crucial source of knowledge in a constructive manner, we need to survey these meta-models (together with the elements they contain, the purposes for which they were built and the extent of their validations), analyse them, and classify them under different perspectives. Nonetheless, both a survey and a framework that categorise and provide a general rationale of all the meta-models described in literature are still absent. 

Indeed, while several SRLs and surveys on Model Driven Engineering (MDE), and Model Driven Architecture (MDA) exist (see e.g., \cite{Silva2015ModeldrivenEA, Santiago2012ModelDrivenEA, Loniewski2010ASR, Gonzlez2014FormalVO, Nguyen:2015:ESR:2831506.2831600}), a SRL on the different types of meta-models available in the field of BPM is still lacking. 
This lack of descriptive categorisation has several negative consequences: the first, and obvious one, is the lack of a comprehensive and easily accessible overview of what has been produced so-far in literature; the second consequence is the danger of over production of \textit{quasi same} meta-models across the BPM community;
 a third consequence is the lack of a framework to categorise and compare the different proposals, which can act as a comprehensive common ground where to place new  proposals of meta-models; and, finally, an investigation is missing on the characteristics, strengths and limits of the current meta-models, so as to identify gaps that may originate further investigations. 

Thus, we identify the needs for this SLR in (i) the lack of reference language independent business process meta-models and, instead, the presence of a number of different, partially overlapping, and sometimes conflicting, meta-models in the literature; (ii) the absence of a systematic study on the meta-models developed in the field of BPM, as well as, (iii) the need of a framework (and its dimensions) to categorise and compare meta-models and the elements they contain.  

\subsection{The research questions}
\label{sec:RQ}

Starting from the needs identified and described in the previous section we have formulated four research questions that motivate and guide our investigation. They are:
\quad \begin{description}
	\item[\textbf{RQ1}.] What types of business process meta-models are being proposed in literature and how can we characterise and categorise them?
	\item[\textbf{RQ2}.] What are the business process elements recurring across business process meta-models?
	\item[\textbf{RQ3}.] What is the role of a business process meta-model? 
	\item[\textbf{RQ4}.] Are the proposed business process meta-models evaluated? How? 
\end{description}

\rqone focuses on the differences among BPML meta-models and aims at investigating them. It also aims at identifying which are the relevant characteristics that meta-models share or in which they differ.

\rqtwo is devoted to the identification of the elements and components of business processes that
 occur in meta-models. Besides providing a photograph of the different components, this research question aims at investigating which are the elements of a business process that are (more) often represented in meta-models and whether these elements correspond to the ones that often occur in the definition of a business process. 

\rqthree is devoted to the identification and classification of the purpose for which the meta-models were introduced / used in the investigated works.

Finally, \rqfour aims at investigating the way the proposed meta-models are evaluated. This question lies on two different motivations. The first, obvious one is to map how meta-models of business processes are evaluated; the second is to assess the  importance provided to the evaluation of meta-models in different studies and to identify suggestions for possible evaluation methodologies. Indeed in literature, there is a lack of guidelines and evaluation criteria for the development of meta-models in the area of business process models and this can hamper their perceived usefulness and (practical) adoption. 

\subsection{The review  protocol}
\label{sec:protocol}

The review protocol was designed around four main phases: (i) data source and strategy; (ii) inclusion and exclusion criteria; (iii) development of the quality assessment; and finally (iv) data extraction strategy and analysis.

\subsubsection{Data source and strategy}
\label{sec:queries}
In this phase, we planned the paper repositories and search queries to be used in our SLR. We decided to perform two different types of searches. First, we decided to target paper repositories, and retrieve papers by means of keyword-based queries. Second, we decided to target proceedings of relevant conferences.

The paper repositories we decided to target are academic peer reviewed paper repositories: DBLP\footnote{\url{https://dblp.uni-trier.de/}}, Scopus\footnote{\url{https://www.scopus.com/search/form.uri?display=basic}}, and Web of Science\footnote{\url{https://login.webofknowledge.com/error/Error?PathInfo=2F&Error=IPError}} (WoS). 
Scopus and WoS were considered because of their extensive coverage on well established scientific literature, especially journal papers. DBLP was included because of its extensive coverage of papers in computer science, including papers published in peer reviewed conference and workshop proceedings.     
To formulate the keyword-based query, we queried the three paper repositories 
considering several combinations of keywords (e.g., process, process model, business process, business process modelling languages, petrinet, petri net, Declare, workflow, meta-model, metamodel) connected by the logical operators AND and OR. The result was the adoption of the query:
\begin{equation}\label{eq:query}
\begin{split}
	& {\texttt{(metamodel OR meta-model) AND}} \\
	& {\texttt{(business process OR process model OR}}\\
	& {\texttt{petrinet OR petri-net OR workflow OR Declare)}}
\end{split}
\end{equation}

whose actual implementation in the syntax of the three repositories is shown in Table~\ref{table:query}. 
 
\begin{table}[t]
\centering
\scalebox{0.9}{
\begin{tabular}{lm{11cm}}
\toprule
 Scopus& \begin{minipage}{9.5cm}
  		\texttt{(``metamodel'' OR ``meta-model'') AND (``business process'' OR ``process model'' OR ``petrinet'' OR ``petri-net'' OR ``workflow'' OR ``Declare'')}
	  \end{minipage} \\ 
 \addlinespace[7pt]
 DBLP& \begin{minipage}{6.5cm}
 		\texttt{metamodel\textbar meta-model AND \\ business process\textbar process model \textbar \\ petrinet\textbar petri-net\textbar workflow\textbar Declare}  
	\end{minipage}\\
 \addlinespace[7pt]
 WoS& \begin{minipage}{11.5cm}
 		\texttt{((TS =``metamodel'' OR TS=``meta-model'') AND \\
		(TS=``business process'' OR TS=``process model'' OR \\
		TS=``petrinet'' OR TS=``petri-net'' OR \\
		TS=``workflow'' OR TS=''Declare'')) AND LANGUAGE:(English)}
	\end{minipage}\\ \bottomrule
\end{tabular}
}
\caption{Key-words on Scopus, DBLP, and WoS.}
\label{table:query}
\end{table}

The proceedings we included in the data sources are the ones of the two reference conference venues in the BPM research area, namely the \textit{Business Process Management} (BPM) conference series\footnote{\url{https://link.springer.com/conference/bpm}} and the  \textit{Conference on Advanced Information Systems Engineering} (CAiSE) series\footnote{\url{https://link.springer.com/conference/caise}}. 


\subsubsection{Inclusion and Exclusion criteria} 
\label{sub:inclusion_and_exclusion_criteria}

The next step of the protocol was to define some relevant criteria in order to evaluate the appropriateness of the papers returned as query results for this study and thus filtering them. The outcome of this step is a set of \emph{primary studies}, that is the set of individual studies contributing to the systematic literature review, which, in turn, represents a form of \emph{secondary study}~\cite{kitchenham2004procedures}.

\begin{table}
\centering
\scalebox{0.9}{
\begin{tabular}{lp{11cm}}
\toprule
IC 1:& The paper proposes a meta-model of business processes or BPMLs.\\
IC 2:& The meta-model is either originally developed or originally adapted by the authors.\\
IC 3:& The paper focuses mainly / exclusively on business process aspects. \\  \midrule
EC 1:& The paper is not available.\\
EC 2:& The paper is duplicate.\\
EC 3:& The paper is not in English.\\
EC 4:& The paper does not belong to the BPM field.\\
EC 5:& The paper does not mainly consider the business process view, but rather it is 
focused on organisational$\backslash$entrepreneurial$\backslash$software engineering aspects without touching the business process level.\\
EC 6:& The paper either was not under peer-review, or it is a technical report.\\
EC 7:& The paper is almost the ``same copy'' of others of the same author(s).\\
EC 8:& The paper either does not include a wide analysis of related works or is not clearly positioned in the state of the art.\\
EC 9:& The paper is not long enough to present a complete meta-model.\\
\bottomrule
\end{tabular}}
\caption{Inclusion and Exclusion Criteria.}
\label{table:iec}
\end{table}

Inclusion (IC) and Exclusion (EC) criteria are reported in Table~\ref{table:iec}. In order to be included, papers had to satisfy all inclusion criteria IC 1 -- IC 3. Moreover, they were excluded if they did satisfy at least one of the exclusion criteria between EC 1 and EC 9.
Basically, all these inclusion and exclusion criteria focus on removing duplicate, incomplete or not scientifically valid papers, papers that are not in English, or refer to the primary criterion of this review, i.e.,
 the paper should be in the BPM field and has to present a meta-model of business processes. Moreover, to maintain the SLR focused, and the amount of papers manageable, we restricted ourselves only to papers where the business process aspect is the main/exclusive focus of the paper, thus excluding papers mainly focused on enterprise, service or software engineering (meta-)models.  
In this phase we decided not to consider ECs limiting the papers selection according to the date of publication. The reason for this choice lies in the fact that this is the first SLR in this field. Thus, we felt we had to consider the maximum number of papers available in literature.

\subsubsection{Quality assessment}
\label{sec:qa}

The four quality assessment criteria we planned and used in this SLR are: 
\begin{itemize}
	\itemsep=-\parsep
\item QA1: Is a well-defined methodology used?
\item QA2: Is the study clearly positioned within the state-of-the-art landscape?
\item QA3: Is the goal of the study elucidated?
\item QA4: Was the study evaluated/validated?
\end{itemize}
We decided to use QA1--QA4 to mark papers with three possible scores: \textit{Yes (Y)}, \textit{No (N)}, and \textit{Partially (P)}, weighted 1, 0 and 0.5 respectively. A description of how QA1--QA4 were used to mark papers can be find in the file called ``quality\_assessment.pdf" at the following link: \url{https://drive.google.com/drive/folders/1xylwk-0OTH8-qWuC2TyvSqB3gbUFpggW}.

\subsubsection{Data extraction strategy and analysis}
\label{sec:selectionStrategy}

\begin{figure}[b]
  \centering
    \includegraphics[width=.9\textwidth]{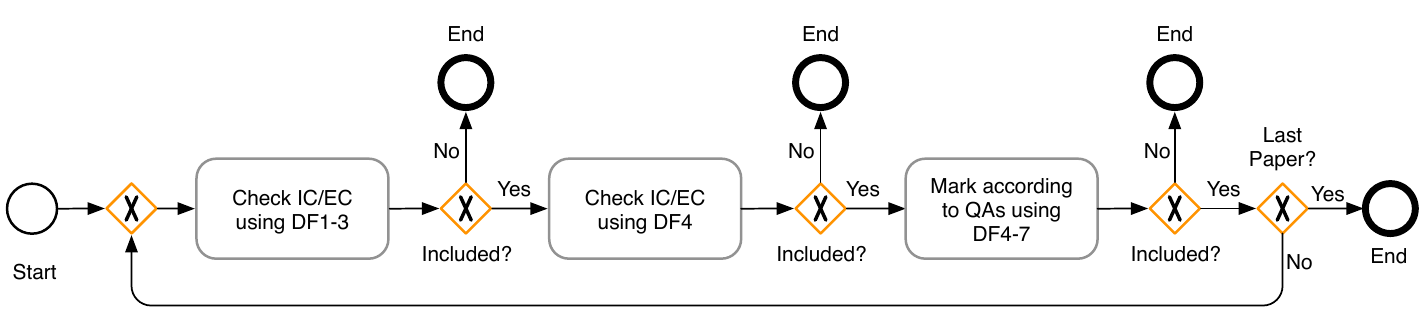}
  \caption{The selection of the primary studies.}
  \label{fig:IMG_SelectionStrategy}
\end{figure}

Within this step, we planned both the data fields of the papers that were used in order to select the primary studies, i.e., the studies analysed for addressing the research questions of the review, and the exact procedure for selecting them. 
The outcome of this phase was a list of data fields to be used for the selection process, contained in Table~\ref{table:DFs}, and the procedure to select the primary studies, graphically illustrated in Figure~\ref{fig:IMG_SelectionStrategy}. The procedure is composed of three steps: (1) all candidate papers must be evaluated against the IC/EC exploiting just title, authors, abstract, and keywords (when present); (2) the the IC/EC are evaluated more carefully on remaining papers using the entire content of the paper; (3) the candidate primary studies are marked using the four quality assessment criteria described in Section~\ref{sec:qa} and are included in the primary studies whenever they score at least 2.5 out of the maximum possible score of 4. 

\begin{table}
\centering
\scalebox{1}{
\begin{tabular}{l@{\qquad\qquad}l}
\toprule
DF1: title and authors & DF2: abstract\\
DF3: keywords & DF4: content\\
DF5: related works & DF7: citations\\
\multicolumn{2}{l}{DF6: publication type (journal, conference, workshop, and book)}\\
\bottomrule
\end{tabular}}
\caption{Data Fields.}
\label{table:DFs}
\end{table}

\section{Extraction of the primary studies}
\label{sec:results}

This section briefly describes the extraction of the primary studies according to the plan presented in the previous section, and the outcomes of each single step in the process (data search, application of inclusion and exclusion criteria, and quality assessment). While the planning was conducted in March 2018, the extraction of the primary studies and of the data necessary for answering the research questions was performed between April 2018 and April 2020.

\paragraph{Conducting the data search}

\begin{table}
\centering
\scalebox{1}{

\begin{tabular}{lS[table-number-alignment = right]S[table-number-alignment = right]cccS[table-number-alignment = right]}
 	  & {Query} &  \multicolumn{1}{c}{No} &  &  &  & \multicolumn{1}{c}{In Primary}\\
  {Source} & {Results} &  \multicolumn{1}{c}{Collections}  & & & &  \multicolumn{1}{c}{Studies}  \\
 \cmidrule{1-3} \cmidrule{7-7} 
 Scopus & 3895  & 3567 &    & & & 60 \\
 WoS    & 542   & 540  &    & & & 21 \\
 DBLP 	& 63 	& 62   &    & & & 5\\
 \cmidrule{1-3} \cmidrule{7-7} 
 CAiSE & & 1065 & No & After & After & \multicolumn{1}{c}{4}\\
 BPM & & 452 & Duplicates & IC/EC & QA & \multicolumn{1}{c}{0}\\
 \midrule
Total  &  & 5686  & 5177 &  67 & 65 & 
\end{tabular}
}
\caption{Query results and selection of Primary Studies.}
\label{table:searchresults}
\end{table}

Papers were selected using the keyword-based queries. 
 Their numbers are reported in the first column of Table~\ref{table:searchresults}. 4500 papers were returned (3895 from Scopus, 542 from WoS, and 63 from DBLP), which were reduced to 4169 after the deletion of collections (e.g., entire proceedings) which were not considered as a single item in this survey. All 452 papers from the BPM conferences (starting from 2003 to 2018) and all 1065 papers published in the CAiSE conferences (starting from 1990 to 2018) were also included in the initial set of papers to be considered\footnote{We have not considered papers related to keynotes speeches and tutorials from both the BPM and CAiSE proceedings.}.
The resulting 5686 papers were pruned from duplicates (papers appearing more than once in the same data source or in at least two data sources) and retracted articles, thus reducing the total number of candidates to 5177.
\footnote{Details of all the retrieved papers, and of the ones removed in each step can be found in the CSV (Comma Separated Values) files
 accessible starting from the folder at \url{https://drive.google.com/drive/folders/1xylwk-0OTH8-qWuC2TyvSqB3gbUFpggW}.}

\paragraph{Applying the inclusion and exclusion criteria}
 
The next step was to apply the IC/EC described in Table~\ref{table:iec} to the 5177
 papers that constitute our starting data collection. To this aim, we manually inspected the papers.
 As a result of this step, 67 papers were retained. These 67 papers were then evaluated according to the quality criteria.

\paragraph{Performing the quality assessment}
\label{QAr}

\begin{figure}[h]
  \centering
    \includegraphics[width=.5\textwidth]{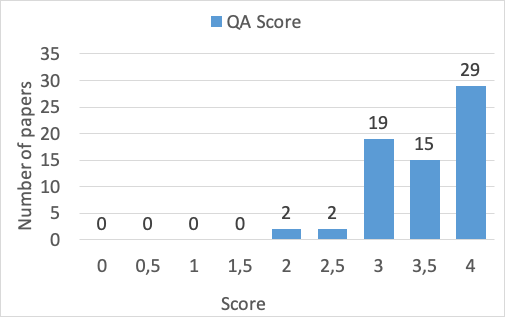}
  \caption{Results grouped by paper.}
  \label{QAresults:papers}
\end{figure}

\begin{figure}[h]
  \centering
	\begin{tabular}{rrS[table-number-alignment = right]rr}
	\toprule
	    & \multicolumn{1}{c}{\textbf{QA1}} & \multicolumn{1}{c}{\textbf{QA2}} & \multicolumn{1}{c}{\textbf{QA3}} & \multicolumn{1}{c}{\textbf{QA4}} \\
	\cmidrule{2-5}
	Yes & 66 & 59 & 67 & 31 \\
	Partially & 1 & 7 & 0 & 15\\
	No & 0 & 1 & 0 & 21\\
	\bottomrule
	\end{tabular}
	    \caption{Results grouped by QA.}
	    \label{QAresults:qa}
\end{figure}

A summary of the quality assessment evaluation is reported in Figures~\ref{QAresults:papers} and~\ref{QAresults:qa}. Most of the papers scored high on most of the questions, with 29 papers scoring \emph{Yes} in all four questions, 15 papers scoring 3.5 in total, 19 papers scoring 3 in total, 
2 papers scoring 2.5 and only the remaining two papers scoring 2 (see Figure~\ref{QAresults:papers}). 
All the \emph{No} answers - except one - concerned the evaluation, where 21 papers out of 67 had a negative score as they did not report any evaluation. The other \emph{No} answer concerns the clarity of the positioning of the paper in the state-of-the-art (see Figure~\ref{QAresults:qa}).

As a result of the quality assessment, the two papers scoring lower than the threshold defined in Section~\ref{sec:selectionStrategy} (2.5) were removed and only 65 papers were retained. These papers constitute our primary studies and are listed in Table~\ref{table:primarystudieslist} classified as workshop, conference (symposium), and journal publications. Their distribution per year is reported in Figure~\ref{fig:Pictures_PSdistributionyear2}, while their venue of publication is reported in Appendix~\ref{app:sourcelist}.

\newcommand{\gap}{.2cm}
 \begin{table}[tbp]
 \centerline{
 \scalebox{.8}{
 \begin{tabular}{llll}
 \toprule
  Year & Workshop Reference & Conference Reference & Journal Reference\\ 
  \midrule
	2001 &					& 			& \citet{10.1016/S0169-023X(01)00034-9} \\
  	2002 &					& \citet{DBLP:conf/caise/SoderstromAJPW02} & \\
			 & 					& \citet{DBLP:conf/adbis/EderG02} & \\
			 & 					& \citet{Groznic2002} & \\[\gap]
  2003 &  					& 						& \citet{DBLP:journals/jkm/PapavassiliouM03}\\
	     &        		& 						& \citet{DBLP:journals/dss/KwanB03} \\ [\gap]
  2004 & 					& \citet{DBLP:conf/adbis/MomotkoS04} & \\
	     & 					& \citet{DBLP:conf/esws/GasevicD04} & \\ [\gap]
  2005 &  \citet{thom2005improving}	& \citet{DBLP:conf/caise/RussellAHE05} & \\ 
  	   & 	\citet{DBLP:conf/er/ListK05} & \citet{Albert2005}	& \\
			 & \citet{DBLP:conf/ihis/HamriBB05} & 	& \\[\gap]
  2006 &  					& \citet{DBLP:conf/sac/ListK06} & \citet{DBLP:journals/ejis/Rittgen06}\\
  	   & 				    & \citet{DBLP:conf/caise/WeigandJABEI06} & \\ 
			 & 						& \citet{DBLP:conf/www/HallerOK06} & \\[\gap]
  2007 & \citet{DBLP:conf/otm/GoedertierV07}& \citet{DBLP:conf/iceis/KorherrL07}	& \citet{DBLP:journals/ijbpim/AxenathKR07}\\
  	   & 					  & \citet{DBLP:conf/iceis/CombemaleCGTV07}	& \citet{Farrell2006FormalisingW}\\ [\gap]
  2008 & \citet{DBLP:conf/edoc/MilanovicGW08}	&   \citet{DBLP:conf/er/RosaDHMG08} & \citet{DBLP:journals/ijbpim/RosemannRF08}\\ [\gap]
  2009 & \citet{DBLP:conf/caise/BessaiN09} & \citet{DBLP:conf/wecwis/ReddingDHI09} & \\[\gap]
  2010 & \citet{Bouchbout2010} & \citet{DBLP:conf/dexa/NicolaMPS10} \\ 
       & 					& \citet{DBLP:conf/icis/HuaZS10} & \\
       & 					& \citet{DBLP:conf/sac/SantosAG10} & \\
			 & 				  & \citet{DBLP:conf/ifip8-1/GaoK10} & \\[\gap]
  2011 & \citet{DBLP:conf/caise/HeidariLK10} 	& \citet{DBLP:conf/edoc/BruningG11} & \citet{DBLP:journals/infsof/StrembeckM11}\\
       & \citet{DBLP:conf/bpmn/Natschlager11} & \citet{DBLP:conf/hicss/WeissW11a} \\
			 &					& \citet{Stroppi2011} &  \\ [\gap]
  2012 & \citet{DBLP:conf/widm/BernardiCLM12}	& \citet{Mahdi2012} & \\
			 & 					& \citet{DBLP:conf/hicss/FriedenstabJMM12} & \\[\gap]
  2013 &					& \citet{DBLP:conf/iceis/BouneffaA13} & \citet{DBLP:journals/jodsn/CherfiAC13}\\
  	   & 					& \citet{DBLP:conf/wecwis/HeidariLBB13}	& \citet{DBLP:journals/is/DamaggioHV13}\\
       & 					& \citet{DBLP:conf/iceis/RamdoyalPDSLJ13} & \\[\gap]
  2014 & \citet{DBLP:conf/wise/KunchalaYY14} & \citet{DBLP:conf/caise/RuizCEFP14}	\\
			 &				& \citet{DBLP:conf/bibm/BraunSBE14} & \\[\gap] 
  2015 &          & \citet{DBLP:conf/es/SprovieriV15} & \citet{martins2015business}\\
	     &          & \citet{DBLP:conf/es/FanesiCH15} & \\
			 & 					& \citet{Thabet2015} & \\[\gap]
  2016 & \citet{DBLP:conf/modellierung/JannaberKRTDB16}	 & \citet{DBLP:conf/isda/HassenTG16} & \citet{DBLP:journals/infsof/ArevaloCRD16}\\
	   & 					& \citet{krumeich2016modeling} & \\
		 & 					& \citet{DBLP:conf/iiwas/YahyaBMB16} & \\
		 &					& \citet{DBLP:conf/kes/OualiMB16} & \\
		 &					& \citet{DBLP:conf/wecwis/StratigakiNLA16} & \\[\gap]
 2017 & 					& \citet{hassen2017extending} & \citet{DBLP:journals/eswa/MertensGP17}\\
	   & 					& \citet{DBLP:conf/wi/DorndorferS17}  \\ [\gap]
	2018 &        & \citet{DBLP:conf/iceis/AhnKK18} \\ [\gap]
	2019 &        &         & \citet{DBLP:journals/access/AmjadAABRN18} \\
\bottomrule	   
\end{tabular}
}}
    \caption{The Primary Studies.}
    \label{table:primarystudieslist}
\end{table}

\begin{figure}[h]
  \centerline{	
    \includegraphics[width=.6\textwidth]{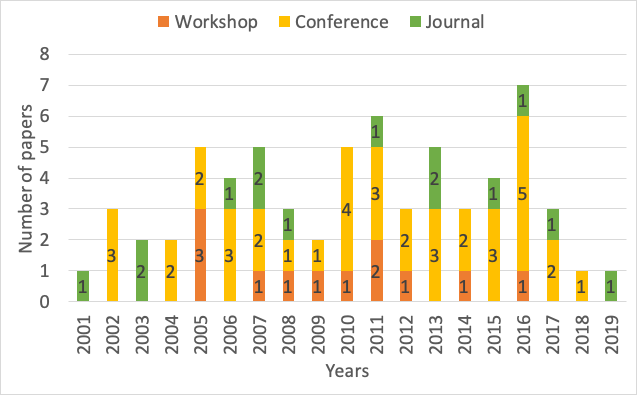}
	}
  \caption{A distribution of the primary studies along the years.}
  \label{fig:Pictures_PSdistributionyear2}
\end{figure}

As summarised in the last column of Table~\ref{table:searchresults}, 60 of these 65 papers were extracted (at least) from Scopus, 21 (at least) from WoS, 5 (at least) from DBLP, 4 (at least) from CAiSE.

\section{A Brief Summary of the Primary Studies}
\label{sec:summaryofpapers}
In this section we report a concise description of the papers included in the primary studies. For the sake of readability the summary is structured in two parts. First, we describe all papers presenting a meta-model of a generic business process model with no reference to a specific business process modelling language; second, we describe papers which provide meta-models specific to a given business process modelling language.

\subsection{Primary Studies proposing language-independent meta-models} 
\label{sec:general_meta_models}


A first group of papers (\cite{DBLP:conf/wecwis/HeidariLBB13,DBLP:conf/sac/ListK06,DBLP:conf/caise/SoderstromAJPW02,DBLP:conf/wise/KunchalaYY14}) proposes general meta-models that are used to compare, integrate, translate, or evaluate specific business process modelling languages. In particular, \citet{DBLP:conf/wecwis/HeidariLBB13} propose a meta-model abstraction obtained by integrating concepts from different business process modelling languages that is then used to classify and compare different process modelling languages.
Similar efforts are contained in \cite{DBLP:conf/sac/ListK06, DBLP:conf/caise/SoderstromAJPW02} where rich meta-models that encompass the typical behavioural perspective are provided. While these four papers consider business process modelling languages that follow the so-called procedural approach to business process modelling, the work of \citet{DBLP:conf/wise/KunchalaYY14}  exploits the meta-model contained in the BALSA framework~\cite{Bhattacharya09adata-centric} to provide a comparative review of modelling languages, with special emphasis to the ones that follow the so-called artifact-centric modelling approach to business process. 

Similarly to this group of works, \citet{DBLP:conf/adbis/EderG02} and \citet{Albert2005} propose general workflow meta-models for workflow transformation and composition. Specifically, \citet{DBLP:conf/adbis/EderG02} present a meta-model for block structured workflow models that supports the hierarchical composition (and hence the reuse) of complex activities. The notions introduced in the meta-model are then used by the authors for providing a definition of workflow equivalence and a set of basic transformations preserving the semantics of the workflows. \citet{Albert2005} present a workflow meta-model, which includes part of the UML2 specification related to activity diagrams, used for applying configuration techniques to address the automatic workflow composition problem.

A third set of papers (\cite{DBLP:conf/www/HallerOK06,Groznic2002}) proposes meta-models that aim at providing a mapping towards or a translation into implementation-oriented representations (e.g., choreography interfaces or information systems). 
\citet{DBLP:conf/www/HallerOK06} propose an ontology (\emph{m3po}) that aims at connecting workflows and choreography descriptions. The usage of the ontology for extracting an Abstract BPEL model from an IBM Websphere MQ Workflow model is reported as initial validation of the proposed approach.

Another group of papers (\cite{DBLP:conf/caise/HeidariLK10,DBLP:conf/icis/HuaZS10,DBLP:journals/jodsn/CherfiAC13,thom2005improving}) exploits meta-models to foster the quality of business process models. 
\citet{DBLP:conf/caise/HeidariLK10} introduce a general meta-model of a business process, inspired by a set of specific business process modelling languages, and they enrich it
with different quality related information connected to the relevant modelling constructs. Instead, \citet{DBLP:conf/icis/HuaZS10} and \citet{DBLP:journals/jodsn/CherfiAC13} address  the issue of improving the quality of a business process model by exploiting domain knowledge. Both works aim at integrating the meta-models of domain ontologies to the one of a business process. Moreover, \citet{DBLP:conf/icis/HuaZS10} present a domain ontology based procedure towards business process modelling, while \citet{DBLP:journals/jodsn/CherfiAC13} describe a set of general mappings between the two meta-models, and their instantiation in the Object Constraint Language (OCL). Finally, \citet{thom2005improving} aim at fostering the quality of business processes via the usage of business (sub)process patterns, and introduce a meta-model called ``Transactional Metamodel of business process'', whose aim is to guide the definition and usage of patterns based on task flow descriptions as well as organisational structural aspects. The meta-model also supports the generation of patterns through BPEL4WS, Business Process Execution Language for Web Services. 

Moving from quality to flexibility, the work of \citet{DBLP:journals/ijbpim/RosemannRF08} proposes the use of a meta-model to represent relevant contextual information in business processes as a way to improve their flexibility and  adaptability. \citet{DBLP:conf/iceis/BouneffaA13} focus on the management of change in software applications based on business process models.
The work proposes a meta-model of business processes extended with a taxonomy of business process change operations. The paper also provides an implementation using a software change management platform based on a set of the Eclipse Workbench plug-ins. 
\citet{DBLP:conf/wecwis/ReddingDHI09} state that in order to be flexible, processes have to be modelled as interacting business objects rather
than as chains of activities. To this aim they introduce a meta-model for business process modelling based on business objects that is used in the paper on a real-life case study to design a number of human service delivery processes.

A further set of works (\cite{DBLP:conf/caise/RussellAHE05,DBLP:conf/iiwas/YahyaBMB16,DBLP:conf/caise/BessaiN09,DBLP:conf/iceis/RamdoyalPDSLJ13,10.1016/S0169-023X(01)00034-9,DBLP:conf/caise/WeigandJABEI06}) focuses on the extension of process model meta-models with specific notions and concepts.

Organisational related aspects represent a rather frequent type of process meta-model extension~\cite{DBLP:conf/caise/RussellAHE05,DBLP:conf/iiwas/YahyaBMB16,DBLP:conf/caise/BessaiN09,DBLP:conf/iceis/RamdoyalPDSLJ13,10.1016/S0169-023X(01)00034-9}. 
\citet{DBLP:conf/caise/RussellAHE05}, in order to present modelling patterns that involve the control flow dimension together with the resource one,  present a rich description of workflow and resource concepts, including the relations that hold among them, which constitute a de-facto meta-model. 
\citet{DBLP:conf/iiwas/YahyaBMB16} extend a Business Process meta-model with concepts related to the social dimension, thus building a Social Business Process (SBP) meta-model. Moreover, the authors also propose an extension to the BPMN meta-model in order to capture SBP requirements.  
The actor and resource focus of the meta-models reported in~\cite{DBLP:conf/caise/BessaiN09} 
 is instead mainly directed towards resource allocation purposes. While in the first paper, the meta-model is used for an actor-driven approach which aims at taking into account the additional work generated by the environment, in the latter, the meta-model is used in order to solve optimization questions on the resource allocation. 
Focusing on workflows, an enhanced workflow meta-model able to capture the resource dimension is presented and evaluated on an industrial case study in~\cite{DBLP:conf/iceis/RamdoyalPDSLJ13}. 
In~\cite{10.1016/S0169-023X(01)00034-9}, instead, a Team-enabled Workflow Reference Model, a meta-model that extends the state-of-the-art organisational meta-model with teams, is presented. The resulting meta-model is then used for making workflow management systems team-enabled.

\citet{DBLP:conf/caise/WeigandJABEI06} extend the behavioural view of business processes with the notion of value. In particular their work
 provides an analysis on the notion of value (objects) in the context of business processes, or, more specifically, in the context of the activities involved when transferring value objects between business actors.

Another group of works (\cite{DBLP:journals/dss/KwanB03,DBLP:conf/kes/OualiMB16,DBLP:conf/ifip8-1/GaoK10}) go beyond the typical behavioural view of business process modelling languages, covering the other typical dimensions of business processes.
In detail, \citet{DBLP:journals/dss/KwanB03} propose KnowledgeScope, a knowledge management system that addresses the problem of the non-use of knowledge repositories by leveraging a process meta-model for organizing knowledge around the organizational processes in which the knowledge is created. The process elements and their relationships within the meta-model are not limited to the behavioural perspective but also covers the functional, organisational and informational perspective. KnowledgeScope has been implemented and applied in a real life organization.
Similarly, \citet{DBLP:conf/kes/OualiMB16} and \citet{DBLP:conf/ifip8-1/GaoK10}, which aim at fostering process model sharing and reuse, take into account different perspectives, going beyond the typical behavioural one. In particular, \citet{DBLP:conf/kes/OualiMB16} propose a Multidimensional Business Knowledge Modeling approach relying on the identification of the perspectives characterizing the business knowledge and on the construction of a Business Knowledge Ontology (BKO). \citet{DBLP:conf/ifip8-1/GaoK10} introduce instead the Business Process Characterizing Model (BPCM) meta-model, whose elements are not limited to the behavioural ones. The meta-model defines a common structure for the models and is used in the BCPM repository architecture, which is proposed to foster the process model reuse.
Finally, the work in~\citet{DBLP:conf/edoc/BruningG11} presents a meta-model expressed in the UML and OCL languages oriented to the representation of workflows in a declarative manner. The meta-model covers the behavioural, data-related and organisational aspects of workflows.

A final set of papers provides meta-models that cover further dimensions related to business processes, including knowledge-intensive processes (\cite{DBLP:conf/isda/HassenTG16,DBLP:journals/jkm/PapavassiliouM03}), perspectives involving different layers (\cite{DBLP:journals/ijbpim/AxenathKR07,martins2015business}), as well as the interoperability across organizations (\cite{Bouchbout2010,DBLP:conf/ihis/HamriBB05}). 

In particular, the works by \citet{DBLP:conf/isda/HassenTG16,DBLP:journals/jkm/PapavassiliouM03} focuses on knowledge-intensive business processes. \citet{DBLP:conf/isda/HassenTG16} propose a rich meta-model for business processes which covers the functional, organisational, behavioural, informational, intentional and knowledge dimensions. The meta-model is then applied to a real world medical scenario. Also focusing on knowledge-intensive processes, \citet{DBLP:journals/jkm/PapavassiliouM03} aim at integrating knowledge and process management. The work
presents an approach for the integration of knowledge tasks and knowledge objects in business process models fostered by meta-models.

\citet{DBLP:journals/ijbpim/AxenathKR07} propose AMFIBIA, a business process meta-model covering a wide set of static and dynamic aspects of processes together with their interactions. The meta-model is also used to realise a formalism-independent workflow engine with the same name. 
The work in \citet{martins2015business} presents a meta-model that enriches the typical constructs of business process modelling languages with layers and concepts coming from work practice information, and in particular with service, structure, and activity related concepts. 

Finally, a special role is played by works focusing on aspects related to interoperability across organizations~\cite{Bouchbout2010} or among workflows~\cite{DBLP:conf/ihis/HamriBB05}. \citet{Bouchbout2010} present a meta-model that captures a wide range of Interorganisational Business Process (IOBP) elements, i.e., a meta-model that aims at characterising the interaction between organizations through business processes. 
\citet{DBLP:conf/ihis/HamriBB05} present, instead, a generic meta-model that workflow models need to share. This meta-model is part of an architecture, with a meta-model, model and data layer, that aims at enabling the semantic, syntactic and control interoperability among workflow models.

\subsection{Primary Studies proposing language specific meta-models} 
\label{language_specifMM}


This group of papers can be further divided into papers focusing on well-known business process modelling languages (\cite{Stroppi2011,DBLP:journals/infsof/ArevaloCRD16,hassen2017extending,DBLP:conf/bibm/BraunSBE14,DBLP:conf/wi/DorndorferS17,DBLP:conf/edoc/MilanovicGW08,DBLP:conf/hicss/FriedenstabJMM12,DBLP:conf/bpmn/Natschlager11,DBLP:conf/es/FanesiCH15,DBLP:conf/iceis/KorherrL07,DBLP:conf/modellierung/JannaberKRTDB16,krumeich2016modeling,DBLP:conf/sac/SantosAG10,DBLP:journals/access/AmjadAABRN18,DBLP:conf/er/RosaDHMG08,DBLP:conf/iceis/CombemaleCGTV07,DBLP:conf/esws/GasevicD04,Mahdi2012,Thabet2015,DBLP:conf/es/SprovieriV15,DBLP:journals/is/DamaggioHV13,Farrell2006FormalisingW,DBLP:journals/infsof/StrembeckM11,DBLP:conf/caise/RuizCEFP14,DBLP:conf/iceis/AhnKK18,DBLP:conf/widm/BernardiCLM12,DBLP:conf/wecwis/StratigakiNLA16}) or on a novel language proposed in the paper, or in related papers by the same authors, together with its meta-model (\cite{DBLP:conf/adbis/MomotkoS04,DBLP:conf/dexa/NicolaMPS10,DBLP:conf/hicss/WeissW11a,DBLP:conf/otm/GoedertierV07,DBLP:journals/eswa/MertensGP17}).  

\paragraph{BPMN}
Within the group tailoring popular business process modelling languages, a conspicuous number of papers (\cite{Stroppi2011,DBLP:journals/infsof/ArevaloCRD16, hassen2017extending,DBLP:conf/bibm/BraunSBE14,DBLP:conf/wi/DorndorferS17,DBLP:conf/edoc/MilanovicGW08,DBLP:conf/hicss/FriedenstabJMM12,DBLP:conf/bpmn/Natschlager11,DBLP:conf/es/FanesiCH15})
 refer to meta-models for BPMN or for BPMN extensions. 
\citet{Stroppi2011} propose an extension to the BPMN 2.0 meta-model and notation related to the resource perspective and, in particular, to the distribution of work among human resources. In detail, the extension concerns the resource structure, the work distribution and the authorization aspect. The extension has then been validated against the Workflow Resource Patterns~\cite{DBLP:journals/itm/Muehlen04}, a set of requirements related to the resource perspective.
The work of \citet{DBLP:journals/infsof/ArevaloCRD16} proposes to extend BPMN 2.0 with a time related perspective. In particular it provides a  taxonomy of declarative rules based on (1) a BPMN meta-model extension that incorporates the time dimension, and (2) an OCL (Object Constraint Language) formalisation of the time related declarative rules. An example of application of how the proposed time-based extension can help in the extraction of business processes from legacy databases is also provided.  

\citet{hassen2017extending} present BPMN4KM, a BPMN 2.0 extension that
focuses on the knowledge dimension for Sensitive Business Processes (SBPs). The extension is done by exploiting BPM4KI, a business process independent generic meta-model common to current BPM formalisms. 
\citet{DBLP:conf/wi/DorndorferS17} provide a BPMN 2.0 meta-model extension tailored to the business processes executed in mobile contexts, that is, business processes supported by mobile devices/applications. 
The work in \citet{DBLP:conf/bibm/BraunSBE14} introduces BPMN4CP, an extension of BPMN2.0 for the representation of clinical pathways. The application of BPMN4CP is demonstrated through an example process related to the wisdom tooth treatment.
\citet{DBLP:conf/edoc/MilanovicGW08} proposes rBPMN (Rule-based BPMN), a BPMN extension that provides support for rules by means of a new type of rule gateway. The rBPMN meta-model integrates the BPMN and the R2ML~\cite{r2ml} meta-models by using the Model Driven Engineering principles. 
In \citet{DBLP:conf/hicss/FriedenstabJMM12}, BPMN is extended with relevant concepts related to Business Activity Monitoring (BAM), i.e., the real-time monitoring and control of business processes. The paper reports the extended BPMN meta-model and an extended BPMN notation enriched with graphical elements representing BAM concepts. The proposed extension has finally been applied to a demo scenario.

Differently from the previous works introducing extensions to BPMN 2.0, the work in \citet{DBLP:conf/bpmn/Natschlager11} provides a BPMN meta-model in the form of formal ontologies. Specifically, the authors introduce a formal ontology for BPMN 2.0, together with some examples of usage, such as a knowledge base and as syntax checker for BPMN 2.0 models.
The last paper of this group (\citet{DBLP:conf/es/FanesiCH15}) also proposes a BPMN ontology; however, they focus on a unified ontology that allows for describing business process meta-model, process model and execution data. The proposed ontology is based on OWL-FA, a multilayer OWL-DL ontology in which every class is tagged with the layer it belongs to. OWL-FA enables a semantic representation of processes and data in an ontology, while providing decidable reasoners.

\paragraph{EPC}
The second most popular language in our primary studies is EPC (Event-Driven Process Chain)~\cite{EPC-Davis}, which is investigated in \cite{DBLP:conf/iceis/KorherrL07,DBLP:conf/modellierung/JannaberKRTDB16,krumeich2016modeling,DBLP:conf/sac/SantosAG10,DBLP:journals/access/AmjadAABRN18,DBLP:conf/er/RosaDHMG08}. 

Actually, \citet{DBLP:conf/iceis/KorherrL07} address both BPMN and EPC. In fact, the paper provides an extension of both the EPC and the BPMN meta-models that adds the concepts of process goals and performance measures. 
In \citet{DBLP:conf/modellierung/JannaberKRTDB16}, the problem of the lack of an official standard EPC meta-model is faced. To fill this gap, the authors propose an integrated EPC meta-model built by taking into account the literature concerning the EPC language and its variants.
\citet{krumeich2016modeling} aim at modelling complex event patterns in EPC and automatically transform them into an executable Event Pattern Language (EPL). The modelling of the complex event patterns exploits an extension of the EPC meta-model proposed in the paper, together with a modelling technique incorporated in the ARIS Business Process Analysis Platform. 
ARIS EPCs are also the focus of the work of~\citet{DBLP:conf/sac/SantosAG10}. This paper presents an 
ontological analysis of the EPC business process modelling notation supported in the ARIS Toolset. The ontological analysis makes use of the Unified Foundational Ontology (UFO)~\cite{Guizzardi-Wagner-UFO} and of a further meta-model of the ARIS Method, introduced by the same authors in \cite{IJBPIM-ARIS}. It provides a precise real-world semantics for business process models represented through EPCs as supported by the ARIS Toolset.
\citet{DBLP:journals/access/AmjadAABRN18} propose UMLPACE (Unified Modeling Language Profile for Atomic and Complex Events in EPC), a UML profile for representing simple and complex EPC patterns. To this aim, the EPC meta-model is first extended with six complex patterns and, then, the concepts of UML activity diagram adapted for representing the EPC meta-model. UMLPACE has a twofold advantage: (i) it allows developers, who are familiar with UML, to model EPC processes; and (ii) through a transformation engine able to transform UMLPACE source models into timed automata, it allows for the verification of complex business requirements. UMLPACE has been validated in two case studies.
The last paper in this group is proposed by~\citet{DBLP:conf/er/RosaDHMG08}, who provide an extensive meta-model for configurable processes with advanced features for capturing resources involved in the performance of tasks  as well as flow of data and physical artefacts. While being potentially applicable to other notations, the meta-model is defined as an extension of EPCs. 

\paragraph{Petri nets and PNML}
Another important group of works focuses on meta-models related to (Coloured) Petri nets (\cite{DBLP:conf/iceis/CombemaleCGTV07,DBLP:conf/esws/GasevicD04,Mahdi2012}) or to the Petri net Markup Language (PNML), i.e., the Petri net standardised language (\cite{Thabet2015}).

In~\citet{DBLP:conf/iceis/CombemaleCGTV07} a Petri net meta-model is used in the proposed approach, which aims at specifying and verifying temporal properties over process models. In detail, in the proposed approach: (i) a temporal extension of OCL for expressing temporal properties (TOCL) is introduced; (ii) a mapping for translating TOCL constraints into LTL formulae is presented; (iii) a mapping between the meta-model of a source language (in the paper SIMPLEPDL, an experimental language, is used for this aim) and the Petri net meta-model is established for translating process models from the source language to Petri nets; (iv) existing model checker tools are used to verify the LTL formulae over the Petri nets.
\citet{DBLP:conf/esws/GasevicD04} present a Petri net ontology for the semantic description of Petri net concepts and their relationships. The ontology is built starting from a Petri net UML meta-model, that, in turn, has been modeled by taking into account Petri net specifications, meta-models, ontologies, and syntax available in the literature.
The work in \citet{Mahdi2012} focuses on Coloured Petri nets, extensions of Petri nets that allow tokens to have a data value. The authors define a business pattern and a Coloured Petri net meta-model and propose an approach based on graph transformation that leverages the meta-model mapping for the automated transformation of business patterns into Coloured Petri nets. Examples have been reported in the paper to illustrate more in detail the proposed approach.
Finally, \citet{Thabet2015} propose to extend Petri nets with the cost perspective. To this aim, they extend the PNML meta-model with cost-related entities and relations and propose an approach for Petri net cost extension based on event logs annotated with cost information.

\paragraph{Artefact-centric languages}
Two further papers (\cite{DBLP:conf/es/SprovieriV15,DBLP:journals/is/DamaggioHV13}) refer to the so-called artefact-centric approach to business process modelling. 

\citet{DBLP:conf/es/SprovieriV15} propose an algorithm to support the run-time planning of partly structured parts of a business process modelled in the CMMN (Case Management Model Notation) modelling language. The run-time planning is used to find an appropriate sequence of tasks. The selection and specification of tasks is supported by an extension of the CMMN meta-model. 
The work of~\citet{DBLP:journals/is/DamaggioHV13} is instead focused on the Guard-Stage-Milestone (GSM) formalism. The GSM meta-model is introduced together with three different, and provably equivalent, formulations of the GSM operational semantics. 

\paragraph{UML}
Another group of papers (\cite{DBLP:journals/ejis/Rittgen06,DBLP:conf/er/ListK05,DBLP:journals/infsof/StrembeckM11}) focuses on the UML language.
Specifically, the purpose of \citet{DBLP:journals/ejis/Rittgen06} is mediating the action view of business process modelling and the reaction view of information systems. To this aim, the author leverages DEMO (Dynamic Essential Modelling of Organization) and UML meta-models for providing a mapping between the two languages. DEMO, which is an enterprise-modelling approach for managing organizations and representing business processes, offers the action view, while UML represents the reaction view.
The work of \citet{DBLP:journals/infsof/StrembeckM11} aims at combining business processes and role-based access control (RBAC) models. To reach this goal they propose a general and language-independent formal meta-model for process-related RBAC models and they also instantiate this meta-model as an extension for UML2 activity models.  
Finally, the work in \citet{DBLP:conf/er/ListK05} proposes a UML 2 profile for Business Process Modelling that provides two complementary perspectives: the business perspective (including aspects like goals, customers and measures) and the control flow perspective. The UML profile is indeed the mechanism that UML 2 provides for extending and adapting its meta-model to a specific area of application. With the Business Process Modelling profile, the UML meta-model is extended so as to take into account the business perspective.

\paragraph{Other procedural languages}
The last papers addressing existing specific BPM procedural languages are \cite{Farrell2006FormalisingW}, 
\cite{DBLP:conf/caise/RuizCEFP14}, and \cite{DBLP:conf/iceis/AhnKK18},
 which focus on YAWL, Communication Analysis, and XPDL, respectively.

\citet{Farrell2006FormalisingW} provide a formal specification of business process workflows. The authors start by representing business processes in terms of a meta-model called \emph{Liesbet}, which
 is based on YAWL patterns, and then formally characterise Liesbet using Milner's Calculus of Communicating Systems (CCS).
\citet{DBLP:conf/caise/RuizCEFP14} aim to align and integrate a goal-oriented modelling language, namely i*,  and a business process-oriented modelling language, namely Communication Analysis, through a reference ontology called FRISCO. The authors also illustrate guidelines for a top-down usage of the two languages, and a tool to support the approach.
The work in \citet{DBLP:conf/iceis/AhnKK18} presents the XPDL (XML Process Definition Language) 2.1 meta-model and an abstracted version of the same meta-model that represents hierarchical relations among entity types in XPDL. Starting from the relational rules of the meta-model, a process entity
hierarchy generator is proposed with the aim of making it easier to perform analytics of hierarchical structures of XPDL process entities.

\paragraph{Declarative languages}
A special group of papers (\cite{DBLP:conf/widm/BernardiCLM12,DBLP:conf/wecwis/StratigakiNLA16}) focuses on meta-models of business process declarative languages.

In \citet{DBLP:conf/widm/BernardiCLM12} a meta-model of Declare is leveraged in the proposed M3D (Model Driven Development with Declare) tool. M3D is a tool for developing Web Applications integrating the Declare meta-model with the three Model Driven Engineering (MDE) meta-models used to represent the structure of information, service and presentation layers of a Web Application.
\citet{DBLP:conf/wecwis/StratigakiNLA16} propose to extend the meta-model of the Compliance Request Language (CLR) \cite{DBLP:journals/sosym/ElgammalTHP16}, an abstract pattern-based language with a LTL (Linear Temporal Logic)/MTL (Metric Temporal Logic) semantics, with a \emph{definition pattern} entity. The extended CRL meta-model is used to show how to describe compliance rules in CRL. Moreover, a graphical view for representing CRL constraints is proposed by the authors. 

\paragraph{New languages}
We conclude with 5 papers presenting meta-models that are used to introduce novel business process modelling languages (\cite{DBLP:conf/adbis/MomotkoS04,DBLP:conf/dexa/NicolaMPS10,DBLP:conf/hicss/WeissW11a,DBLP:conf/otm/GoedertierV07,DBLP:journals/eswa/MertensGP17}). 

\citet{DBLP:conf/adbis/MomotkoS04} present the business process query language BPQL. To do this, a meta-model of workflows is introduced to represent the workflow model upon which BPQL has to operate. 
\citet{DBLP:conf/dexa/NicolaMPS10} present a platform for business process modelling and verification. The platform is centred around the logic-based language BPAL (Business Process Abstract Language), introduced by some of the
 authors in \cite{DBLP:conf/iicai/NicolaLM07}. BPAL mainly focuses on the control flow perspective of business processes and is illustrated in the paper together with its meta-model.  
\citet{DBLP:conf/hicss/WeissW11a} introduce the semantic process modelling language SBPML, together with its meta-model. SBPML is a domain-specific language tailored to the financial sector and thus, its meta-model contains domain-independent elements as well as domain-specific ones describing financial processes related elements in all the process, organisation, data object and resource views. 
EM-BrA$^{2}$CE, an extension of the Semantics of Business Vocabulary and Business Rules (SBVR), is introduced in~\citet{DBLP:conf/otm/GoedertierV07}. The EM-BrA$^{2}$CE vocabulary allows for referring to the state of a business process, thus enabling to describe the business rules that govern business processes. 
Finally, the last work of this group (\citet{DBLP:journals/eswa/MertensGP17}) presents the meta-model of \emph{DeciClare}, a new business process declarative language that is able to capture information related to the typical four perspectives of business processes, i.e., control-flow, functional, data and resource. 
The effectiveness of the proposed language has been evaluated by conducting interviews with domain experts.

\section{Answering the research questions}
\label{sec:RRQs}

This section presents the answers of all the research questions introduced in Section~\ref{sec:RQ}. 

\subsection{Answering \rqone}
\label{subsec:RQ1}

In answering \rqone we aim at investigating the characteristics of the meta-models introduced in the literature and whether there is a way to categorise them. 
This question could have several answers, depending on the perspective exploited to look at the meta-models. In this paper, we answer \rqone in two different steps. 

The first characterisation we make of the papers is based on their relationship with specific modelling languages or paradigms.\footnote{This characterisation is roughly the one that we have exploited in reporting the concise description of the primary studies provided in Section~\ref{sec:summaryofpapers}. Even if the categorisation of the primary studies in different groups was obtained when answering \rqone, and will therefore be discussed here, we decided to exploit it also in Section~\ref{sec:summaryofpapers} for the sake of presentation.}  
Indeed, by looking at the meta-models of the \psnumber primary studies, we can observe that they can be divided 
in two different categories: the first one, called \gen, which contains general meta-models of business processes that are not related to any concrete business process modelling language; the second, hereafter named \lsp, which contains meta-models of concrete business process modelling languages. In turn, \lsp can be divided in two sub-categories: the first one, called \oldl, contains meta-models of an existing well-established business process modelling language, while the second one, hereafter called \newl, contains meta-models of new modelling language proposed in the very same paper, or by the same author in closely related papers.  

\begin{table}[ht]
\centering
\scalebox{0.8}{
\begin{tabular}{ll}
\toprule
  Category & Primary studies  \\ \midrule
	 \multirow{2}{*}{\gen} & \cite{DBLP:conf/caise/SoderstromAJPW02,DBLP:conf/wecwis/HeidariLBB13,DBLP:conf/sac/ListK06,DBLP:conf/wise/KunchalaYY14,DBLP:conf/caise/HeidariLK10,DBLP:conf/isda/HassenTG16,martins2015business,DBLP:conf/edoc/BruningG11,DBLP:conf/caise/WeigandJABEI06,DBLP:journals/ijbpim/AxenathKR07,DBLP:journals/jkm/PapavassiliouM03,DBLP:conf/caise/RussellAHE05,DBLP:conf/icis/HuaZS10,DBLP:journals/jodsn/CherfiAC13,thom2005improving,DBLP:journals/ijbpim/RosemannRF08,DBLP:conf/iceis/BouneffaA13}\\ 
	& \cite{DBLP:conf/adbis/EderG02,Albert2005,DBLP:conf/www/HallerOK06,Groznic2002,DBLP:conf/wecwis/ReddingDHI09,DBLP:conf/iiwas/YahyaBMB16,DBLP:conf/caise/BessaiN09,DBLP:conf/iceis/RamdoyalPDSLJ13,10.1016/S0169-023X(01)00034-9,DBLP:journals/dss/KwanB03,DBLP:conf/kes/OualiMB16,DBLP:conf/ifip8-1/GaoK10,Bouchbout2010,DBLP:conf/ihis/HamriBB05}\\
	\lsp \\
	\quad \multirow{2}{*}{\oldl} & \cite{DBLP:journals/infsof/ArevaloCRD16, hassen2017extending,DBLP:conf/wi/DorndorferS17,DBLP:conf/bpmn/Natschlager11,DBLP:conf/iceis/KorherrL07,krumeich2016modeling,DBLP:conf/sac/SantosAG10,DBLP:conf/er/RosaDHMG08,DBLP:conf/es/SprovieriV15,DBLP:journals/is/DamaggioHV13,Farrell2006FormalisingW,DBLP:journals/infsof/StrembeckM11,DBLP:conf/caise/RuizCEFP14} \cite{Stroppi2011,DBLP:conf/bibm/BraunSBE14,DBLP:conf/edoc/MilanovicGW08,DBLP:conf/hicss/FriedenstabJMM12}\\
& \cite{DBLP:conf/es/FanesiCH15,DBLP:conf/modellierung/JannaberKRTDB16,DBLP:journals/access/AmjadAABRN18,DBLP:conf/iceis/CombemaleCGTV07,DBLP:conf/esws/GasevicD04,Mahdi2012,Thabet2015,DBLP:journals/ejis/Rittgen06,DBLP:conf/er/ListK05,DBLP:conf/iceis/AhnKK18,DBLP:conf/widm/BernardiCLM12,DBLP:conf/wecwis/StratigakiNLA16}\\	
	\quad \newl & \cite{DBLP:conf/adbis/MomotkoS04,DBLP:conf/dexa/NicolaMPS10,DBLP:conf/hicss/WeissW11a} \cite{DBLP:conf/otm/GoedertierV07,DBLP:journals/eswa/MertensGP17}\\
\bottomrule
\end{tabular}}
\caption{A first characterisation of meta-models.}
\label{table:GLOtable}
\end{table}

Table~\ref{table:GLOtable} provides the list of these categories (where indentation is used to indicate subclasses), together with a classification of the primary studies w.r.t. the categories just introduced. In short, 31 papers present meta-models that are independent from any specific modelling language\footnote{We have classified the paper in \citet{DBLP:conf/iiwas/YahyaBMB16} as independent from business process languages, although it reports both a generic business process meta-model extension and a specific extension for BPMN.}, while 34 papers belong to the language specific class \lsp. Of the latter, the biggest group is the one describing meta-models of existing business process modelling languages (29 papers), while the remaining 5 papers present meta-models of newly proposed business process modelling languages.

By looking at the primary studies we noticed further characteristics of the meta-models, ranging from the scope of the meta-model, to the type of language used to express it, to the tool support provided in the approach.\footnote{Note that, in answering \rqone we do not take into account the process model elements described by the meta-models (e.g., whether they enable to describe roles, goals, artefacts and so on). This is due to the fact that we have a specific research question (\rqtwo) devoted to investigate what is described by the meta-models.} This second set of categories we extracted from the primary studies takes into account:
\begin {enumerate}[(i)]
\item whether the meta-model is described in terms of a formal or of a semi-formal language:
\begin{itemize}
	\itemsep=-\parsep
\item Formal (\frm): the meta-model is described by means of a formal language;
\item Semi-formal (\semifrm): the meta-model is described through a semi-formal representation (e.g., the meta-model is described as a UML diagram);
\end{itemize}
\item whether the meta-model considers only the process model dimension or both the process model and execution dimension:
\begin{itemize}
	\itemsep=-\parsep
\item Meta-models of models (\model): the meta-model considers (only) the process model dimension;
\item Meta-models of executions and models  (\modexe): the meta-model considers both the process execution and the process model dimensions; 
\end{itemize}
\item whether the meta-model adheres to a procedural, declarative or hybrid view of business processes:
\begin{itemize}
	\itemsep=-\parsep
\item Procedural (\proc): the meta-model adheres to a procedural view of business processes; 
\item Declarative (\dec): the meta-model adheres to a declarative view of business processes; 
\item Hybrid (\hyb): the meta-model adheres to a hybrid (including both procedural and declarative) view of business processes;
\end{itemize}
\item whether the meta-model adheres to an activity- or artefact-centric view of business processes:
\begin{itemize}
	\itemsep=-\parsep
\item Activity-centric (\act): the meta-model adheres to an activity-centric view of business processes; 
\item Artefact-centric (\art): the meta-model adheres to an artefact-centric view of business processes; 
\end{itemize} 
\item whether the meta-model is domain dependent or independent:
\begin{itemize}
	\itemsep=-\parsep
\item Domain-dependent (\dm): The meta-model is domain dependent;
\item Domain-independent (\dmi): The meta-model is not domain dependent;
\end{itemize}
\item whether the meta-model is (somehow) evaluated or not:
\begin{itemize}
	\itemsep=-\parsep
\item Evaluated (\eval): The meta-model is (somehow) evaluated;
\item Non-evaluated (\neval): The meta-model is not evaluated.
\end{itemize}

\end {enumerate}

\begin{table}[ht]
\centering

\scalebox{.78}{
\begin{tabular}{lll}
\toprule
  Class & Primary studies & \#  \\ \midrule    
 \frm & \cite{DBLP:journals/infsof/ArevaloCRD16,DBLP:conf/bpmn/Natschlager11,DBLP:conf/er/RosaDHMG08,DBLP:conf/es/SprovieriV15,DBLP:journals/is/DamaggioHV13,Farrell2006FormalisingW,DBLP:journals/infsof/StrembeckM11, DBLP:conf/dexa/NicolaMPS10,DBLP:conf/es/FanesiCH15,DBLP:conf/esws/GasevicD04,DBLP:conf/kes/OualiMB16,DBLP:conf/iceis/CombemaleCGTV07,10.1016/S0169-023X(01)00034-9,DBLP:conf/er/ListK05,Albert2005} & 15\\ 
\semifrm & all except \cite{DBLP:journals/infsof/ArevaloCRD16,DBLP:conf/bpmn/Natschlager11,DBLP:conf/er/RosaDHMG08,DBLP:conf/es/SprovieriV15,DBLP:journals/is/DamaggioHV13,Farrell2006FormalisingW,DBLP:journals/infsof/StrembeckM11, DBLP:conf/dexa/NicolaMPS10,DBLP:conf/es/FanesiCH15,DBLP:conf/esws/GasevicD04,DBLP:conf/kes/OualiMB16,DBLP:conf/iceis/CombemaleCGTV07,10.1016/S0169-023X(01)00034-9,DBLP:conf/er/ListK05,Albert2005} & 40 \\ \hline
  \multirow{2}{*}{\model} & \cite{DBLP:conf/wecwis/HeidariLBB13,DBLP:conf/sac/ListK06, DBLP:conf/isda/HassenTG16,  martins2015business,DBLP:conf/caise/RuizCEFP14, DBLP:conf/caise/WeigandJABEI06,DBLP:conf/icis/HuaZS10, DBLP:journals/jodsn/CherfiAC13,DBLP:journals/ijbpim/RosemannRF08,DBLP:journals/infsof/ArevaloCRD16,  hassen2017extending,  DBLP:conf/wi/DorndorferS17,DBLP:conf/bpmn/Natschlager11, DBLP:conf/iceis/KorherrL07,DBLP:conf/sac/SantosAG10,DBLP:conf/hicss/WeissW11a} & \multirow{2}{*}{40} \\ 
&\cite{thom2005improving, DBLP:conf/es/SprovieriV15,DBLP:journals/eswa/MertensGP17,DBLP:conf/iiwas/YahyaBMB16,Mahdi2012,DBLP:journals/access/AmjadAABRN18,Bouchbout2010,DBLP:conf/iceis/AhnKK18,DBLP:conf/modellierung/JannaberKRTDB16,DBLP:journals/dss/KwanB03,DBLP:conf/widm/BernardiCLM12,Stroppi2011,DBLP:journals/ejis/Rittgen06,DBLP:conf/kes/OualiMB16,DBLP:conf/ifip8-1/GaoK10,DBLP:conf/er/ListK05,DBLP:conf/caise/BessaiN09,DBLP:conf/ihis/HamriBB05,DBLP:conf/bibm/BraunSBE14,DBLP:conf/wecwis/StratigakiNLA16,Groznic2002,DBLP:conf/edoc/MilanovicGW08,Albert2005,DBLP:conf/hicss/FriedenstabJMM12} & \\ 
	\modexe & \cite{DBLP:conf/caise/SoderstromAJPW02,DBLP:conf/wise/KunchalaYY14, DBLP:conf/caise/HeidariLK10,DBLP:conf/edoc/BruningG11,  DBLP:journals/ijbpim/AxenathKR07, DBLP:conf/caise/RussellAHE05,DBLP:conf/iceis/BouneffaA13, krumeich2016modeling,DBLP:conf/er/RosaDHMG08,Farrell2006FormalisingW,  DBLP:journals/infsof/StrembeckM11,DBLP:conf/dexa/NicolaMPS10,DBLP:journals/jkm/PapavassiliouM03,DBLP:journals/is/DamaggioHV13,DBLP:conf/adbis/MomotkoS04,DBLP:conf/es/FanesiCH15,DBLP:conf/esws/GasevicD04,DBLP:conf/wecwis/ReddingDHI09,DBLP:conf/iceis/RamdoyalPDSLJ13,DBLP:conf/adbis/EderG02,DBLP:conf/iceis/CombemaleCGTV07,10.1016/S0169-023X(01)00034-9,DBLP:conf/www/HallerOK06,Thabet2015,DBLP:conf/otm/GoedertierV07} &  25 \\ \hline
	\proc & all, except	\cite{DBLP:conf/wise/KunchalaYY14,DBLP:conf/edoc/BruningG11,DBLP:conf/dexa/NicolaMPS10, DBLP:journals/infsof/ArevaloCRD16,DBLP:journals/is/DamaggioHV13,DBLP:conf/es/SprovieriV15,DBLP:conf/caise/WeigandJABEI06,DBLP:journals/eswa/MertensGP17,DBLP:conf/widm/BernardiCLM12,DBLP:conf/wecwis/ReddingDHI09,DBLP:conf/wecwis/StratigakiNLA16,DBLP:conf/otm/GoedertierV07,DBLP:conf/iceis/CombemaleCGTV07,Groznic2002,DBLP:conf/edoc/MilanovicGW08} & 50\\
\dec & \cite{DBLP:conf/wise/KunchalaYY14,DBLP:conf/edoc/BruningG11,DBLP:conf/dexa/NicolaMPS10, DBLP:journals/infsof/ArevaloCRD16,DBLP:journals/is/DamaggioHV13,DBLP:conf/es/SprovieriV15,DBLP:journals/eswa/MertensGP17,DBLP:conf/widm/BernardiCLM12,DBLP:conf/wecwis/ReddingDHI09,DBLP:conf/wecwis/StratigakiNLA16,DBLP:conf/otm/GoedertierV07} & 11\\
\hyb & \cite{DBLP:conf/iceis/CombemaleCGTV07,Groznic2002,DBLP:conf/edoc/MilanovicGW08} & 3\\ \hline
	\act & all, except \cite{DBLP:conf/wise/KunchalaYY14, DBLP:journals/is/DamaggioHV13,DBLP:conf/es/SprovieriV15,DBLP:conf/caise/WeigandJABEI06,DBLP:conf/wecwis/ReddingDHI09} & 60\\
  \art & \cite{DBLP:conf/wise/KunchalaYY14, DBLP:journals/is/DamaggioHV13,DBLP:conf/es/SprovieriV15,DBLP:conf/wecwis/ReddingDHI09} & 4\\ \hline
  \dm & \cite{DBLP:conf/hicss/WeissW11a,DBLP:conf/wi/DorndorferS17,DBLP:conf/bibm/BraunSBE14} & 3 \\ 
  \dmi & all, except\cite{DBLP:conf/hicss/WeissW11a, DBLP:conf/wi/DorndorferS17,DBLP:conf/bibm/BraunSBE14} & 62 \\ \hline
\multirow{2}{*}{\eval} &  \cite{DBLP:conf/isda/HassenTG16,
	DBLP:conf/wecwis/HeidariLBB13,DBLP:conf/icis/HuaZS10,DBLP:journals/jodsn/CherfiAC13,DBLP:journals/ijbpim/RosemannRF08,DBLP:conf/edoc/BruningG11,DBLP:conf/caise/WeigandJABEI06,DBLP:journals/jkm/PapavassiliouM03,DBLP:journals/infsof/ArevaloCRD16,DBLP:conf/iceis/KorherrL07,krumeich2016modeling,DBLP:journals/infsof/StrembeckM11,DBLP:conf/sac/SantosAG10,DBLP:conf/wi/DorndorferS17,DBLP:conf/hicss/WeissW11a,DBLP:conf/bpmn/Natschlager11} & \multirow{2}{*}{33}\\
	& \cite{DBLP:journals/access/AmjadAABRN18,DBLP:journals/ejis/Rittgen06,DBLP:conf/es/FanesiCH15,DBLP:conf/iiwas/YahyaBMB16,Stroppi2011,DBLP:conf/www/HallerOK06,DBLP:journals/eswa/MertensGP17,Bouchbout2010,DBLP:conf/iceis/AhnKK18,DBLP:conf/widm/BernardiCLM12,DBLP:conf/wecwis/ReddingDHI09,DBLP:conf/iceis/RamdoyalPDSLJ13,10.1016/S0169-023X(01)00034-9,DBLP:conf/er/ListK05,DBLP:conf/ihis/HamriBB05,DBLP:conf/bibm/BraunSBE14,DBLP:conf/hicss/FriedenstabJMM12} & 	\\
\multirow{2}{*}{\neval} & \cite{DBLP:journals/ijbpim/AxenathKR07,DBLP:journals/is/DamaggioHV13,martins2015business,Farrell2006FormalisingW,DBLP:journals/dss/KwanB03,DBLP:conf/wecwis/StratigakiNLA16,DBLP:conf/dexa/NicolaMPS10,DBLP:conf/er/RosaDHMG08,DBLP:conf/adbis/MomotkoS04,DBLP:conf/iceis/BouneffaA13,DBLP:conf/iceis/CombemaleCGTV07,DBLP:conf/caise/SoderstromAJPW02,DBLP:conf/caise/RuizCEFP14,DBLP:conf/caise/RussellAHE05,DBLP:conf/es/SprovieriV15,Mahdi2012} & \multirow{2}{*}{32} \\
& \cite{DBLP:conf/kes/OualiMB16,Thabet2015,Groznic2002,Albert2005,DBLP:conf/ifip8-1/GaoK10,DBLP:conf/edoc/BruningG11,DBLP:conf/adbis/EderG02,DBLP:conf/sac/ListK06,hassen2017extending,DBLP:conf/esws/GasevicD04,DBLP:conf/wise/KunchalaYY14,DBLP:conf/caise/HeidariLK10,DBLP:conf/caise/BessaiN09,DBLP:conf/edoc/MilanovicGW08,DBLP:conf/otm/GoedertierV07,DBLP:conf/modellierung/JannaberKRTDB16} &  \\ 
 \bottomrule
\end{tabular}}
\caption{A second characterisation of meta-models.}
\label{table:Addtable}
\end{table}

Table~\ref{table:Addtable} provides a description of the primary studies w.r.t. the classes introduced above. 
Considering how the meta-model are represented, 15 primary studies provide a formal representation of the meta-model they describe. 40 are focused on the model dimension only,
while 25 take into account both the model and the execution dimension. Concerning the approach towards business process modelling, most primary studies adhere to the traditional procedural and activity-centric based view on business processes (50 and 60 papers respectively), with very few papers taking a declarative (11), hybrid (3), or artefact-centric view (4).\footnote{The work of \citet{DBLP:conf/caise/WeigandJABEI06} appears to provide an original, yet uncommon, ``value centred'' approach towards business process modelling that seems to share some characteristics of artefact-centric declarative approaches. Nonetheless, a classification under the \dec and \art categories was not possible, due to a lack of details. }

Another aspect to be taken into account is the one related to the domain (in)dependency of the meta-model. In our study, only three papers focus on domain-specific business processes, while all the others are domain-dependent. The three domains are the financial sector~\cite{DBLP:conf/hicss/WeissW11a}, a context-sensitive mobile domain~\cite{DBLP:conf/wi/DorndorferS17}, and the clinical domain~\cite{DBLP:conf/bibm/BraunSBE14}. 
Finally, slightly more than 50\% of the meta-models are (somehow) evaluated (\eval), even if the level of evaluation differs greatly among the different papers. This aspect will be better discussed in Section \ref{subsec:RQ4}, when answering \rqfive.\footnote{Please note that QA4 did concern with an evaluation/validation of the study which could encompass the meta-model, while here we refer explicitly to the evaluation of the meta-model.}

\subsection{Answering \rqtwo}
\label{subsec:RQ2}


The aim of \rqtwo is to present an overview of the \textit{elements} involved in the primary studies' meta-models. In answering this question we 
have identified 682 single elements which have been grouped in 15 sets of recurrent constructs across the classes of meta-models. These 15 sets identify macro-elements that appear in the primary studies' meta-models, and are: \emph{activity},  \emph{event}, \emph{event $\grave{a}$-la BPMN}, \emph{state}, \emph{sequence flow}, \emph{rule}, \emph{time}, \emph{data flow}, \emph{data object},  \emph{actor}, \emph{resource}, \emph{capability}, \emph{value}\footnote{Although the explicit element ``value'' only occurs in one of the meta-models of the primary studies \cite{DBLP:conf/caise/WeigandJABEI06} and hence does not explicitly appear among the elements of the \emph{value} group, all the elements included in this group refer to measurable aspects related to the value of a business process.}, \emph{goal}, and \emph{context}.

To focus our analysis on central elements of business processes and exclude variants that were specific to a single meta-model, we decided to concentrate our study only to the 154 (out of 682) elements that are considered in at least two meta-models.
These 154 elements are listed in Table~\ref{elementsRQ2}, together with their corresponding group.\footnote{Note that, five elements belong to more than one set of macro-elements. They are: \texttt{information}, \texttt{position}, \texttt{role}, (software) \texttt{application} and \texttt{process participant}}. Moreover, some elements, such as \texttt{informational resource} and \texttt{non-agentive resource}, sometimes may overlap. For each element we report, in round brackets, the number of primary studies' meta-models in which it occurs, while boldface is used to denote the elements that appear at least in 15\% of the meta-models. In some cases, elements with the same or very similar meaning had different names in the meta-models. To simplify the analysis and the reporting we have classified all the syntactic variants under only one name. 
Finally, for each macro-element we also report in round brackets the number of corresponding elements and the total occurrences of these elements within the meta-models.
The list of the main syntactic variants for each element and the correspondence between each element and the primary studies in which it appears can be found at \url{https://drive.google.com/drive/folders/1xylwk-0OTH8-qWuC2TyvSqB3gbUFpggW}.

Before commenting the table we have to clarify that we are aware of the problems arising from a study in which the information from different sources is blindly brought together. Indeed a problem we had to overcome in extracting the elements and creating the table was the establishment of the semantics of its components (i.e., the labels’ semantics) or, at the very least, the clarification of their intended meaning. In fact, only few authors did include explicit semantics, while for most of the cases it was either lacking or provided in terms of commonsense descriptions. Since our aim here was mainly to survey the elements present in the original meta-models and be faithful to the authors' representations, in order to avoid bias, we relied as much as we could on the descriptions provided by the authors, to use a commonsense semantics of business process (modelling) elements, and to exclude terms for which at least a basic understanding was not acquirable from the paper itself.

\newcommand{\spann}{.9\textwidth}
\begin{table}[htbp]
 \centering
\scalebox{0.8}{
 \begin{tabular}{ll}
 \toprule
 Macro-element & Element\\
 \midrule
 \begin{minipage}{.2\textwidth}
 \textit{activity}\\
 \textbf{(11/122)}
 \end{minipage} &
 \begin{minipage}{\spann}
 \textbf{\texttt{activity}}~\textbf{(43)}, \texttt{function}~\textbf{(6)}, \textbf{\texttt{atomic activity}}~\textbf{(21)}, \textbf{\texttt{compound activity}}~\textbf{(21)}, \textbf{\texttt{transition}}~\textbf{(11)}, 	\texttt{activity instance}~\textbf{(9)}, \texttt{manual activity}~\textbf{(3)}, \texttt{automatic activity}~\textbf{(2)},
 \texttt{collaborative organisational activity}~\textbf{(2)}, \texttt{critical organizational activity}~\textbf{(2)}, \texttt{cancel activity}~\textbf{(2)}
 \end{minipage}
 \\
\cmidrule{2-2}
 \begin{minipage}{.2\textwidth}
 \textit{event}\\
 \textbf{ (7/35)}
 \end{minipage} &
 \begin{minipage}{\spann}
 \textbf{\texttt{event}}~\textbf{(13)}, \texttt{initial node}~\textbf{(5)}, \texttt{final node}~\textbf{(6)}, \texttt{message event}~\textbf{(2)}, \texttt{signal}~\textbf{(2)}, \texttt{occurrence$\slash$event occurrence}~\textbf{(5)}, \texttt{trigger (event)}~\textbf{(2)}
 \end{minipage}
 \\
 \cmidrule{2-2}
 \begin{minipage}{.2\textwidth}
 \textit{event $\grave{a}$-la BPMN}\\
 \textbf{(19/52)}
 \end{minipage} &
 \begin{minipage}{\spann}
 \texttt{event-BPMN}~\textbf{(9)}, \texttt{event sub-process}~\textbf{(2)}, \texttt{throw event}~\textbf{(3)}, \texttt{interrupting}~\textbf{(2)}, \texttt{start event}~\textbf{(4)}, \texttt{intermediate event}~\textbf{(4)}, \texttt{end event}~\textbf{(4)}, \texttt{message event}~\textbf{(2)}, \texttt{link event}~\textbf{(2)}, \texttt{multiple event}~\textbf{(2)}, \texttt{timer event}~\textbf{(2)}, \texttt{escalation event}~\textbf{(2)}, \texttt{error event}~\textbf{(2)}, \texttt{parallel multiple event}~\textbf{(2)}, \texttt{conditional event}~\textbf{(2)}, \texttt{catch event}~\textbf{(2)}, \texttt{event non interrupting}~\textbf{(2)}, \texttt{cancellation event}~\textbf{(2)}, \texttt{conditional event}~\textbf{(2)}
 \end{minipage}
 \\
 \cmidrule{2-2}
 \begin{minipage}{.2\textwidth}
 \textit{state}\\
 \textbf{(12/61)}
 \end{minipage}
  &
 \begin{minipage}{\spann}
 \texttt{state}~\textbf{(5)}, \texttt{event-EPC}~(8), \texttt{atomic event-EPC}~(2) \texttt{complex event-EPC}~(3), \textbf{\texttt{precondition}~\textbf{(17)}}, \texttt{postcondition}~\textbf{(9)}, \texttt{place}~(5), \texttt{state occurrence}~(4), \texttt{event-EPC exclusion}~(2), \texttt{event-EPC sequence}~(2), \texttt{event-EPC cardinality}~(2), \texttt{event-EPC trend}~(2)
 \end{minipage}
 \\
 \cmidrule{2-2}
 \begin{minipage}{.2\textwidth}
 \textit{sequence flow}\\
 \textbf{(23/139)}
 \end{minipage}
 &
 \begin{minipage}{\spann}
 \texttt{sequence}~\textbf{(3)}, \texttt{multimerge}~\textbf{(2)}, \texttt{multi choice}~\textbf{(2)},
 \texttt{syncronisation}~\textbf{(4)}, \textbf{\texttt{connecting object}~\textbf{(13)}},
 \textbf{\texttt{sequence flow}~\textbf{(12)}}, \texttt{merge}~\textbf{(3)}, \texttt{join}~\textbf{(4)}, \texttt{fork}~\textbf{(4)}, \textbf{\texttt{gateway}}~\textbf{(24)}, \texttt{complex} gateway~\textbf{(3)}, 
 \texttt{event-based gateway}~\textbf{(2)}, \textbf{\texttt{parallel gateway}}~\textbf{(16)},
 \textbf{\texttt{inclusive gateway}~\textbf{(13)}}, \textbf{\texttt{exclusive gateway}}~\textbf{(15)}, 
 \texttt{flow operator}~\textbf{(4)}, \texttt{input flow connector}~\textbf{(2)}, \texttt{output flow connector}~\textbf{(2)}, \texttt{unconditional coordination pattern}~\textbf{(2)}, \texttt{existence}~\textbf{(3)}, \texttt{coexistence}~\textbf{(2)}, \texttt{precedence}~\textbf{(2)}, \texttt{absence}~\textbf{(2)}
 \end{minipage}
 \\
\cmidrule{2-2}
 \begin{minipage}{.2\textwidth}
 \textit{rule}\\
 \textbf{(4/9)}
 \end{minipage}
 &
 \begin{minipage}{\spann}
 \texttt{decision rule}~\textbf{(3)}, \texttt{business rule}~\textbf{(2)}, \texttt{assignment} (to an actor)~\textbf{(2)}, \texttt{resource parameter binding}~\textbf{(2)}
 \end{minipage}
 \\
 \cmidrule{2-2}
 \begin{minipage}{.2\textwidth}
 \textit{time}\\
 \textbf{(1/3)}
 \end{minipage} & 
 \begin{minipage}{\spann}
 \texttt{time duration}~\textbf{(3)}
 \end{minipage}
 \\
 \cmidrule{2-2}
 \begin{minipage}{.2\textwidth}
 \textit{data flow} \\
 \textbf{(5/30)}
 \end{minipage} & 
 \begin{minipage}{\spann}
 \textbf{\texttt{message flow}}~\textbf{(10)}, \texttt{data flow}~\textbf{(8)}, 
 \texttt{association}~\textbf{(8)}, \texttt{conversational link}~\textbf{(2)}, \texttt{knowledge flow}~\textbf{(2)}
 \end{minipage}
 \\
 \cmidrule{2-2}
 \begin{minipage}{.2\textwidth}
 \textit{data object}\\
 \textbf{(23/94)}
 \end{minipage} & 
 \begin{minipage}{\spann}
 \textbf{\texttt{artifact}}~\textbf{(15)}, \texttt{physical artifact}~\textbf{(2)},  
 \textbf{\texttt{data object}}~\textbf{(13)}, \texttt{message}~\textbf{(7)}, \texttt{data input}~\textbf{(4)}, \texttt{data output}~\textbf{(4)}, \texttt{conversation}~\textbf{(3)}, 
 \texttt{call conversation}~\textbf{(2)}, \texttt{information}~\textbf{(5)}, \texttt{physical knowledge support}~\textbf{(2)}, 
 \texttt{internal knowledge}~\textbf{(2)}, \texttt{tacit knowledge}~\textbf{(2)}, 
 \texttt{external knowledge}~\textbf{(2)}, \texttt{explicit knowledge}~\textbf{(2)}, 
 \texttt{procedural knowledge}~\textbf{(3)}, \texttt{knowledge}~\textbf{(3)}, 
 \texttt{document}~\textbf{(6)}, \texttt{artifact instance}~\textbf{(2)}, 
 \texttt{data store}~\textbf{(4)}, \texttt{database}~\textbf{(2)}, \texttt{contract}~\textbf{(2)}, \texttt{product}~\textbf{(3)},     \texttt{deliverable}~\textbf{(4)}
 \end{minipage}
 \\
 \cmidrule{2-2}
 \begin{minipage}{.2\textwidth}
 \textit{actor}\\
 \textbf{(18/130)}
 \end{minipage} & 
 \begin{minipage}{\spann}
 \textbf{\texttt{actor}}~\textbf{(25)}, \texttt{collective agent}~\textbf{(5)}, \texttt{actor instance}~\textbf{(4)}, \textbf{\texttt{organisation}~\textbf{(13)}}, \textbf{\texttt{organisation unit}~\textbf{(13)}}, 
 \texttt{human expert}~\textbf{(2)}, \texttt{internal agent}~\textbf{(4)}, \texttt{external agent}~\textbf{(4)}, \texttt{client}~\textbf{(5)}, \texttt{position}~\textbf{(4)}, (software) \texttt{application}~\textbf{(4)}, \textbf{\textbf{\texttt{role}}~\textbf{(22)}}, \texttt{role instance}~\textbf{(2)}, \texttt{process owner}~\textbf{(3)}, \texttt{process participant}~\textbf{(9)}, \texttt{person}~\textbf{(7)}, \texttt{human performer}~\textbf{(2)}, \texttt{organisation structure}~\textbf{(2)}
 \end{minipage}
 \\
 \cmidrule{2-2}
 \begin{minipage}{.2\textwidth} 
 \textit{resource}\\
 \textbf{(16/84)}
 \end{minipage} & 
 \begin{minipage}{\spann}
 \texttt{resource} (agentive)~\textbf{(5)}, \texttt{resource} (non-agentive)~\textbf{(8)}, \texttt{resource} (both agentive and non-agentive) ~\textbf{(4)},  \texttt{resource (unclear)~\textbf{(5)}}, \texttt{material resource}~\textbf{(6)}, \texttt{immaterial resource}~\textbf{(4)}, \texttt{information}~\textbf{(9)}, \texttt{position}~\textbf{(3)}, \textbf{\textbf{\texttt{role}}}~\textbf{(13)}, (software) \texttt{application}~\textbf{(8)},  \texttt{process participant}~\textbf{(5)}, \texttt{software}~\textbf{(3)}, \texttt{service}~\textbf{(4)}, \texttt{resource parameter}~\textbf{(3)}, \texttt{human resource}~\textbf{(2)}, \texttt{non human resource}~\textbf{(2)}
 \end{minipage}
 \\
\cmidrule{2-2}
 \begin{minipage}{.2\textwidth}
 \textit{capability}\\
 \textbf{(3/7)}
 \end{minipage} & 
 \begin{minipage}{\spann}
 \texttt{duties}~\textbf{(2)}, \texttt{skills}~\textbf{(2)}, \texttt{capabilities$\slash$competences}~\textbf{(3)}
 \end{minipage}
 \\
 \cmidrule{2-2}
 \begin{minipage}{.2\textwidth} 
 \textit{value}\\
 \textbf{(5/17)}
 \end{minipage} & 
 \begin{minipage}{\spann}
 \texttt{measure}~\textbf{(5)}, \texttt{cost}~\textbf{(3)}, \texttt{unit} (of measurement)~\textbf{(2)}, \texttt{qualitative measure}~\textbf{(3)}, \texttt{quantitative measure}~\textbf{(4)}
 \end{minipage}
 \\
 \cmidrule{2-2} 
 \begin{minipage}{.2\textwidth}
 \textit{goal} \\
 \textbf{(5/21)}
 \end{minipage} & 
 \begin{minipage}{\spann}
	 \texttt{organisational objective}~\textbf{(4)}, \textbf{\texttt{goal}~\textbf{(11)}}, \texttt{common goal}~\textbf{(2)}, \texttt{soft goal}~\textbf{(2)}, \texttt{hard goal}~\textbf{(2)}
	 \end{minipage}
 \\
 \cmidrule{2-2} 
 \begin{minipage}{.2\textwidth}
  \textit{context}\\
  \textbf{(2/5)}
  \end{minipage} & 
 \texttt{context}~\textbf{(3)}, \texttt{business area}~\textbf{(2)}\\  
 \bottomrule
 \end{tabular}
 }
\caption{Recurring elements in meta-models.}
\label{elementsRQ2}
\end{table} 

As we can see, five sets of macro-elements stand up as distinctive both in terms of different elements and in terms of overall occurrence. 
 They are: \emph{activity}, \emph{sequence flow}, \emph{actor}, \emph{data object} and \emph{resource}.
Perhaps not surprisingly, the most articulate and recurring elements are the ones belonging to the \emph{sequence flow} group, with 23 different elements appearing 139 times in total. Another relevant group is the one of \emph{activity} (122 in total), where we can notice a strong homogeneity in the variety of elements and a big presence of the \texttt{activity} element, which is the most recurring element in all the meta-models.  Interestingly enough, one of the aspects that distinguishes  business process models from other types of processes, that is, the \emph{actor}/organisational aspect, constitutes the  second group in terms of overall occurrence.    
It is indeed well represented in most of the meta-models both in terms of variety of elements (18) and overall presence (130). 
An interesting group is the one of \emph{data object}, where we can notice a detailed description of different types of knowledge (23 in total) that can appear in business process model elements. Similarly interesting is the relevant presence of \emph{resource}-related elements even though the elements of this category presented several challenges which we better describe below.

Further relevant macro-elements are the ones containing the ``event'' related elements. While a single term ``event'' was used in all primary studies it was quite evident from the papers themselves that this term was used with different intuitive meanings: in particular elements labelled as ``event'' where often introduced with a BPMN-like semantics, i.e., ``something that happens during the course of a process''~\cite{model2011notation} or with a EPC-like semantics, i.e., ``describing preconditions and postconditions of functions''~\cite{Mendling2008}. In some residual cases the label ``event'' was used in a way not explicitly connectable to either event $\grave{a}$-la BPMN or event $\grave{a}$-la EPC\footnote{The semantic overloading of the term ``event'' is a well known fact in the BPM community. See e.g., the different definitions of event at \url{https://www.businessprocessglossary.com/11516/event}.}. To make these three different usages clear, we decided to explicitly distinguish between them by classifying the elements into three different macro-elements: event $\grave{a}$-la BPMN as something that happens during the course of a process, event as \emph{state} describing pre- and and post-conditions, and a generic macro-element \emph{event} whose precise meaning we plan to better investigate in the future.

Moving to the less frequent macro-elements, we can note that the key elements of goal (or value) appearing in almost all the modern definitions of business processes (such as the one of Weske provided in Section~\ref{sec:needforSLR}) have a lower (or in some cases just implicit) presence in business process meta-models even though the element \texttt{goal} appears at least in 15\% of the meta-models. Another poorly populated macro-element, composed of elements that have recently gained importance in the BPM community is the one of \emph{rule}. 

While we were able to classify most of the elements in terms of a single macro-element, this operation was not always possible. Indeed there were few cases in which the same element had an unclear, and often overloaded meaning, which - to be faithful to the original meta-models as we said above - we aimed at surveying. Thus, certain elements do appear in more than a single macro-element. This phenomenon involves mainly elements that appear in the macro-element \emph{resource} and in either \emph{actor} or \emph{data object}. This happens because elements such as \texttt{information} or \texttt{process participant} can indeed play different roles in a business process, acting e.g., as an artefact (resp., actor) or as a resource.\footnote{The overlap between the two macro-elements \emph{actor} and \emph{resource} can be due to the use of the term ``human resource'' in organizational sciences, which may lead to classify humans as resources.}

Focusing on single elements, we can notice the big presence of \texttt{activity}, and the fact that most of the meta-models present a distinction between atomic and compound activities. Similarly, for \texttt{gateway} and the different types of gateways. \texttt{actor} and \texttt{role} are two other elements recurring more than 20 times, together with \texttt{organisation}, if we sum it up also with \texttt{organisation unit}. Another frequent term is the one of ``resource''. Similarly to what happened with the term ``event'', we have observed rather different ways of referring to the single element \texttt{resource}: sometimes the term was associated with elements having only agentive characteristics (e.g., agents), other times with elements having only non-agentive characteristics (e.g., artefacts); sometimes a unique term was employed to refer to both resources with agentive and non-agentive characteristics; while sometimes it was also used in a generic or unclear manner. Since for most of the papers it was possible to make this distinction, we decided to make the difference explicit in the table.

As already said, only few meta-models mention goal, rule, and value-related elements. Few are also the meta-models that specify the capabilities of the participants involved in the business processes. Another interesting observation is the fact that \texttt{state} does not appear very often in an explicit manner, but it appears more frequently in the form of pre- and post- conditions, event-EPC and places (of Petri Nets). Some meta-models include in their representation also instance elements, such as \texttt{activity instance} and \texttt{occurrence$\slash$event occurrence} which we decided to document for the sake of completeness of the extraction.  

Focusing on the frequency of the elements in the meta-models, 21 elements appear in at least 15\% of them and are: \texttt{activity}, \texttt{atomic activity}, \texttt{compound activity}, \texttt{transition}, \texttt{event}, \texttt{precondition}, \texttt{connecting object}, \texttt{sequence flow}, \texttt{gateway}, \texttt{parallel gateway}, \texttt{in\-clu\-sive gateway}, \texttt{exclusi\-ve gateway}, \texttt{message flow}, \texttt{artifact}, \texttt{data object}, \texttt{actor}, \texttt{organisation}, \texttt{or\-ga\-nisa\-tion unit}, \texttt{role} (both as an \emph{actor} and as a \emph{resource}), and \texttt{goal}. 
If we increase the threshold to ``appearing in at least 25\% meta-models'', then only 8 elements satisfy it:  \texttt{activity}, \texttt{atomic activity}, \texttt{compound activity}, \texttt{precondition}, \texttt{gateway}, \texttt{parallel gateway}, \texttt{actor} and \texttt{role} as a \emph{actor($\slash$organisation)}. A remark needs to be made here about the element \texttt{resource}. Indeed if we consider it at a very abstract level, ignoring its agentive vs non agentive characterisation, then it also appears in at least 10\% and 25\% of meta-models. Finally, only 1 element (\texttt{activity}) appears in more than 50\% of the studies. 

\subsection{Answering \rqthree}
\label{subsec:RQ3}

The aim of this research question
 is to identify the reason to introduce/use the meta-models in the selected primary studies. Note that the reason to introduce the meta-model does not necessarily coincide with the overall aim of the paper. In fact, the meta-model is often an instrument for reaching a more comprehensive goal rather than being the goal of the paper.   

Table~\ref{table:role} provides a categorisation of the primary studies w.r.t. 21 different purposes we were able to extract from the studies themselves. While extracting the reason to introduce a meta-model is somehow complex, as meta-models can be exploited in several ways, in the table we report only the purposes that were actually deeply described
 and illustrated in the papers, and not, for instance, to the ones that were just mentioned or left for future work and generalisations. 

\begin{table}[ht]
\centering
\scalebox{.8}{
\begin{tabular}{ll}
\toprule
  Class & Primary studies  \\ \midrule    
	\textbf{describe aspects of a business process} & all\\
  \multirow{2}{*}{\textbf{extend a meta-model/ML with new concepts}} & \cite{DBLP:conf/caise/HeidariLK10,DBLP:journals/ijbpim/RosemannRF08,DBLP:conf/iceis/BouneffaA13,DBLP:conf/isda/HassenTG16,martins2015business,DBLP:conf/caise/WeigandJABEI06,DBLP:journals/jkm/PapavassiliouM03,DBLP:conf/caise/RussellAHE05,DBLP:journals/infsof/ArevaloCRD16,hassen2017extending,DBLP:conf/wi/DorndorferS17,DBLP:conf/iceis/KorherrL07,DBLP:conf/er/RosaDHMG08,DBLP:conf/es/SprovieriV15,DBLP:journals/infsof/StrembeckM11,DBLP:conf/caise/RuizCEFP14}\\
  & \cite{DBLP:conf/iiwas/YahyaBMB16,DBLP:conf/iceis/RamdoyalPDSLJ13,DBLP:conf/kes/OualiMB16,10.1016/S0169-023X(01)00034-9,DBLP:conf/caise/BessaiN09,Thabet2015,Groznic2002,Bouchbout2010,Stroppi2011,DBLP:conf/er/ListK05,DBLP:conf/wecwis/StratigakiNLA16,DBLP:conf/otm/GoedertierV07,DBLP:conf/hicss/FriedenstabJMM12}\\
  \textbf{incorporate patterns in meta-model} & \cite{thom2005improving,krumeich2016modeling,DBLP:journals/infsof/StrembeckM11,DBLP:journals/access/AmjadAABRN18}\\
  \textbf{integrate process \& domain ontology} & \cite{DBLP:conf/icis/HuaZS10,DBLP:journals/jodsn/CherfiAC13}\\
  \textbf{support modelling and quality of models} & \cite{DBLP:conf/caise/HeidariLK10,DBLP:conf/icis/HuaZS10,DBLP:journals/jodsn/CherfiAC13,thom2005improving,DBLP:conf/kes/OualiMB16}\\
  \textbf{support flexibility} & \cite{DBLP:conf/wecwis/ReddingDHI09}\\
  \textbf{compare modelling languages} & \cite{DBLP:conf/caise/SoderstromAJPW02, DBLP:conf/wecwis/HeidariLBB13, DBLP:conf/sac/ListK06,DBLP:conf/wise/KunchalaYY14}\\
	  \textbf{map/integrate modelling languages/constructs} & \cite{DBLP:conf/wecwis/HeidariLBB13,DBLP:conf/ihis/HamriBB05,DBLP:conf/www/HallerOK06,DBLP:conf/modellierung/JannaberKRTDB16,DBLP:conf/iceis/CombemaleCGTV07,DBLP:journals/ejis/Rittgen06,DBLP:conf/widm/BernardiCLM12,Mahdi2012}\\
	  \textbf{classify modelling languages} & \cite{DBLP:conf/caise/SoderstromAJPW02}\\
	\textbf{evaluate modelling languages} & \cite{DBLP:conf/sac/ListK06,DBLP:conf/wise/KunchalaYY14}\\
	\textbf{create language independent representation} & \cite{DBLP:conf/wecwis/HeidariLBB13,DBLP:journals/ijbpim/AxenathKR07,DBLP:conf/edoc/BruningG11}\\
	  \textbf{describe a modelling language} & \cite{DBLP:journals/is/DamaggioHV13, Farrell2006FormalisingW}\\
	  \textbf{define a new modelling language} & \cite{DBLP:conf/adbis/MomotkoS04,DBLP:conf/ifip8-1/GaoK10,DBLP:conf/wecwis/ReddingDHI09,DBLP:conf/edoc/MilanovicGW08,DBLP:conf/iceis/CombemaleCGTV07,DBLP:journals/eswa/MertensGP17} \\
	\textbf{clarify semantics of modelling language} & \cite{DBLP:conf/sac/SantosAG10,DBLP:conf/modellierung/JannaberKRTDB16}\\
	\textbf{formal representation of ML} & \cite{DBLP:conf/bpmn/Natschlager11,DBLP:conf/dexa/NicolaMPS10,DBLP:conf/adbis/EderG02}\\
	  \textbf{support reasoning techniques} &  \cite{DBLP:conf/bpmn/Natschlager11,DBLP:conf/dexa/NicolaMPS10,Albert2005,DBLP:conf/esws/GasevicD04,DBLP:conf/es/FanesiCH15,DBLP:conf/iceis/RamdoyalPDSLJ13}\\
  \textbf{extract coreogaphies} & \cite{DBLP:conf/www/HallerOK06}\\
  \textbf{evaluate suitability of a ML for a domain} & \cite{DBLP:conf/hicss/WeissW11a}\\
  \textbf{support extension of a ML to a new domain} & \cite{DBLP:conf/hicss/WeissW11a,DBLP:conf/bibm/BraunSBE14}\\
  \textbf{support workflow-knowledge management} & \cite{DBLP:journals/dss/KwanB03}\\
  \textbf{extract hierarchical information} & \cite{DBLP:conf/iceis/AhnKK18}\\
 \bottomrule
\end{tabular}}
\caption{Why introducing meta-models?}
\label{table:role}
\end{table}

As we can see, all meta-models in our primary studies aim at providing an illustration of aspects of a business process. The second 
 most popular usage of a meta-model in our primary studies was the extension of the meta-model itself, both generic and language specific (i.e., \textit{ML} in the table), with a new concept (29 papers). \cite{DBLP:conf/caise/HeidariLK10} extends it with quality metrics; \cite{DBLP:journals/ijbpim/RosemannRF08,DBLP:conf/wi/DorndorferS17} with a notion of context; \cite{DBLP:conf/iceis/BouneffaA13} with the notion of change and how change relates to business process elements; \cite{DBLP:conf/isda/HassenTG16,DBLP:journals/jkm/PapavassiliouM03,hassen2017extending,DBLP:conf/kes/OualiMB16} with the notion of knowledge, business knowledge, and knowledge-related concepts; \cite{martins2015business} introduces the relation between business processes and daily practices while \cite{Bouchbout2010} extends it with inter-organisational aspects; \cite{DBLP:conf/caise/WeigandJABEI06} extends a business process meta-model with the notion of value, while \cite{Thabet2015} extends it with the notion of cost; \cite{DBLP:conf/caise/RussellAHE05,DBLP:conf/er/RosaDHMG08,DBLP:conf/iceis/RamdoyalPDSLJ13,Stroppi2011} with the notion of resource; \cite{DBLP:conf/er/RosaDHMG08} introduces also a data dimension concerning artefacts and data objects; \cite{DBLP:journals/infsof/ArevaloCRD16} extends a meta-model with the notion of time; \cite{DBLP:conf/es/SprovieriV15,DBLP:conf/iceis/KorherrL07,DBLP:conf/caise/RuizCEFP14} extends it with the notion of goal; \cite{DBLP:conf/iceis/KorherrL07} enriches it also with the notion of performance; the work in \cite{DBLP:journals/infsof/StrembeckM11} extends it with RBAC related concepts (e.g., roles) and also RBAC related workflow patterns; actor and role are also introduced in \cite{DBLP:conf/caise/BessaiN09};  \cite{DBLP:conf/iiwas/YahyaBMB16} extends it with social concepts; \cite{10.1016/S0169-023X(01)00034-9} with the notion of teams; \cite{Groznic2002} with the notion of business rule;  \cite{DBLP:conf/wecwis/StratigakiNLA16} extends the metric temporal logic (MTL) meta-model to support references to compliance sources; and \cite{DBLP:conf/hicss/FriedenstabJMM12} extends a business process meta-model with the notions to support Business Activity Monitoring. Finally two works extend non BPM languages to provide support for the modelling of Business Processes: \cite{DBLP:conf/er/ListK05} does it for UML 2, and \cite{DBLP:conf/otm/GoedertierV07} does it for the SBVR (Semantics of Business Vocabulary and Business Rules) standard. 
Examples of extension of the meta-model are even more present if we consider also the four additional papers that incorporate workflow patterns in the meta-model, the two papers that extend business process meta-models with the ability to connect to domain ontologies, and the two papers that aim to extend modelling languages to a new domain (the financial domain in \cite{DBLP:conf/hicss/WeissW11a} and the clinical domain in \cite{DBLP:conf/bibm/BraunSBE14}). 

Coming to the less frequent usages we can note that 5 papers introduce meta-models to support modelling and the quality of models; 11 papers exploit meta-models for comparing (mapping, integrating, classifying) different modelling languages (including different variants of the same modelling language) and in some cases evaluate them; instead, 13 papers use meta-models for describing an existing modelling language (only), support the definition of a new one, clarify their semantics, or create from them a language independent representation. 
 Another group of papers (7 in total) focuses on the creation of formal representations of meta-models in order to formally specify the behaviour of workflow models, to create an OWL ontology version of the meta-models, or/and exploit different forms of reasoning techniques (e.g., to verify the well formedness of a business model specification or to support workflow composition).

\subsection{Answering \rqfour}
\label{subsec:RQ4}

As already reported in Table~\ref{table:Addtable}, few primary studies present some forms of evaluation of the meta-models they describe. In answering \rqfour we aim at investigating the way these evaluations are carried out. 

Table~\ref{table:evaluation} provides a categorisation of the forms of evaluation we were able to extract from the primary studies.  Given that not many papers provide in depth evaluations, we have listed here also the studies in which use cases are mainly used as illustrative examples of how the meta-model (or the framework that includes the meta-model) can be applied.

\begin{table}[ht]
\centering
\scalebox{.8}{
\begin{tabular}{ll}
\toprule
  Class & Primary studies  \\ \midrule    
  \textbf{Extensive Case Studies} & \cite{DBLP:conf/isda/HassenTG16,DBLP:journals/access/AmjadAABRN18,DBLP:journals/ejis/Rittgen06,DBLP:conf/es/FanesiCH15}\\
	  \textbf{Ontological Analysis} & 	\cite{DBLP:conf/wecwis/HeidariLBB13,DBLP:conf/sac/SantosAG10,DBLP:conf/iiwas/YahyaBMB16}\\
	  \textbf{Comparison with requirements} & \cite{DBLP:conf/wi/DorndorferS17,DBLP:conf/hicss/WeissW11a,Stroppi2011}\\
	  \textbf{Comparison with other meta-models} & \cite{DBLP:journals/access/AmjadAABRN18,DBLP:conf/www/HallerOK06}\\
	  \textbf{Formal properties} & \cite{DBLP:conf/bpmn/Natschlager11} \\
	  \textbf{Conceptual Modeling Quality Framework} & \cite{DBLP:journals/eswa/MertensGP17}\\
	  \textbf{Domain experts} & \cite{DBLP:journals/eswa/MertensGP17}\\
	  \multirow{2}{*}{\textbf{Illustrative examples}} & \cite{DBLP:conf/wecwis/HeidariLBB13,DBLP:conf/icis/HuaZS10,DBLP:journals/jodsn/CherfiAC13,DBLP:journals/ijbpim/RosemannRF08,DBLP:conf/edoc/BruningG11,DBLP:conf/caise/WeigandJABEI06,DBLP:journals/jkm/PapavassiliouM03,DBLP:journals/infsof/ArevaloCRD16,DBLP:conf/iceis/KorherrL07,krumeich2016modeling,DBLP:journals/infsof/StrembeckM11}\\
		& 		\cite{Bouchbout2010,DBLP:conf/iceis/AhnKK18,DBLP:conf/widm/BernardiCLM12,DBLP:conf/wecwis/ReddingDHI09,DBLP:conf/iceis/RamdoyalPDSLJ13,10.1016/S0169-023X(01)00034-9,DBLP:conf/er/ListK05,DBLP:conf/ihis/HamriBB05,DBLP:conf/bibm/BraunSBE14,DBLP:conf/hicss/FriedenstabJMM12}\\
 \bottomrule
\end{tabular}}
\caption{How are the meta-models evaluated?}
\label{table:evaluation}
\end{table}

Overall, only 14 papers present some form of evaluation, while 21 papers only present illustrative examples.
Illustrative examples concern different scenarios and are, thus, the most recurring method to show the applicability of the approach. Despite their popularity, they are not a real form of evaluation and their specific details are therefore omitted in the remaining of the section. 

The only papers that provide real/extensive use cases and exploit them to support precise characteristics of the meta-model based framework are \cite{DBLP:conf/isda/HassenTG16,DBLP:journals/access/AmjadAABRN18,DBLP:journals/ejis/Rittgen06,DBLP:conf/es/FanesiCH15}. \citet{DBLP:conf/isda/HassenTG16} present a real use case taken from a medical domain. Here the aim is to go beyond a mere illustration and to evaluate how the concepts contained in the meta-model can support an understandable, adequate and expressive representation of Sensitive Business Processes. \citet{DBLP:journals/access/AmjadAABRN18} provides two extensive use cases, one taken from at ATM fraud detection and simple transaction process, the other from a patient flow system in an hospital. The case studies help illustrate how the UMLPACE language based on the meta-model is able to capture the extensive requirements of the scenarios. \citet{DBLP:journals/ejis/Rittgen06} provides highlights of an experience from two real projects, one with a logistics provider and another with a large retail chain, where the aim was the modelling of the complex inter-organisational business processes as a basis for their reorganisation. The paper itself reports very little details on the experiments. Nonetheless, the  paper reports excerpts of the models and the findings highlighting how the real experiences were used to assess both the modelling helpfulness of the DEMO language and the mapping framework between DEMO and UML. Finally, \citet{DBLP:conf/es/FanesiCH15} shows the adequacy of the ontology produced in the paper by using it to answer concrete queries over a specific scenario of a student admission process.

A different form of evaluation of the characteristics and quality of the meta-models is provided by \citet{DBLP:conf/wecwis/HeidariLBB13}, \citet{DBLP:conf/iiwas/YahyaBMB16}, and \citet{DBLP:conf/sac/SantosAG10}. These primary studies  exploit an ontological analysis to show how the meta-meta model is successful in expressing concepts taken from upper level ontologies. In the first two papers the upper level ontology used is the Bunge-Wand-Weber (BWW) upper level ontology~\cite{Wand:1990ab}, while in the third it is the UFO upper-level ontology~\cite{Guizzardi-Wagner-UFO}. 

\citet{DBLP:conf/wi/DorndorferS17} provides an evaluation of the extended meta-model by comparing it with the requirements for its development presented at the beginning of the paper. A similar evaluation is provided by \citet{DBLP:conf/hicss/WeissW11a}. In the work of \citet{Stroppi2011} the requirements are instead provided by the Workflow Resource Patterns \cite{DBLP:conf/caise/RussellAHE05}
 which are considered as an abstraction of a set of recurrent requirements regarding the resource perspective.

\citet{DBLP:journals/access/AmjadAABRN18} perform an evaluation by comparing the modelling support for complex events in EPC provided by their meta-model against the one of four other meta-models. Instead \citet{DBLP:conf/www/HallerOK06} aims at validating their proposed multi meta-model process ontology m3po by mapping it to/from other meta-models (IBM Websphere MQ Workflow and Abstract BPEL).\footnote{At the time of writing the links containing the full listing of mappings are not working thus preventing an assessment of the evaluation itself.}

\citet{DBLP:conf/bpmn/Natschlager11} provides an evaluation of the formal ontology in terms of its formal (logic-based) properties of consistency and correctness. 

An extensive evaluation is carried out by \citet{DBLP:journals/eswa/MertensGP17}. In that paper the authors make use of the Conceptual Modeling Quality Framework (CMQF) \cite{Nelson2012} to evaluate the effectiveness of the language with three different quality types: perceived semantic quality, pragmatic quality and language-domain appropriateness, and a realistic arm fracture scenario. These qualities were evaluated within the CMQF by exploiting semi-structured interviews with domain experts.

By looking at these results we can say that a rigorous evaluation of meta-models is often neglected in literature as it reduces, in the majority of cases, to mere illustrative examples. 
Two forms of evaluation stand out from this analysis and can provide the basis for 
guidelines and evaluation criteria for the development of meta-models in the area of business processes. First, an evaluation by means of real use cases: this can help the assessment of the elements contained in the meta-model to support the modelling of real scenarios. Second, an evaluation by means of a comparative analysis. The comparison can involve upper level ontologies, requirements of some form, or other meta-models. These different forms of comparisons can (i) help the assessment of the meta-model w.r.t. needs or conditions that motivated its development; (ii) help the assessment of the meaning and properties of concepts present in the meta-model on the basis of well-known reference elements contained in foundational ontologies; and (iii) help the assessment of the expressivity of meta-model w.r.t. other meta-models. While not recurrent in our primary studies, the evaluation provided by exploiting the CMQF and semi-structured interviews with domain experts is also interesting and should be considered in this sort of studies.    
  
\section{Discussion of Results} 
\label{sec:discussion}

The data presented in Section~\ref{sec:RRQs} enable answering, at least partially, the four research questions presented in Section~\ref{sec:RQ} that were used to shape this SLR. 

Before addressing the research questions in detail, let us comment on the temporal distribution and the distribution by publication type of the primary studies.
Concerning the temporal distribution we can observe that, while we did not pose any temporal restriction towards the data search in Scopus, WoS, and DBLP, and while we also manually evaluated all the CAiSE proceedings from 1990, the first paper included in our primary studies was published 
in 2001. 
Thus, the interest in this area seems to be a recent one\footnote{Interestingly enough, 2003 was the year when the BPM conference series started.} with about 2/3 of the works published between 2007 and 2016.
Overall, the relatively low number of papers identified, and their temporal distribution indicate that this topic is still under-investigated. Also, considering the importance of the topic, and the growing interest in different approaches towards Business Process Modelling (e.g., procedural vs.~declarative or activity-centric vs.~artefact-centric styles of modelling) we were expecting a larger number of publications in the last 10-year period, with a growing trend. Instead, we notice a slight decrease of publications starting from 2017,
which could be related to this lack of a comprehensive common ground where to place new proposals of meta-models.  

Regarding the distribution by publication type, we can notice a reasonable indication of scientific maturity. Indeed, the data contained in Appendix~\ref{app:sourcelist} show a good distribution between journal and conference publications (22\% and 60\% of the total, respectively),
 and - even more important - 1/4
 of the primary studies (25\%) that was published in journals/conferences ranked Q1 or A/A$^*$ according to Scopus/CORE (see details in Appendix~\ref{app:sourcelist}). This number increases to about 43\% if we include also journals/conferences ranked Q2 or B. 
If we restrict only to journal publications, 11 out of 14 (79\% of the total number of journals) belong to the 1st or 2nd Quartile according to the chosen journal ranking.
Not surprisingly, the publication venues mostly refer to the areas of Software Engineering, Conceptual Modelling, and Business Process Management, even though no standard venue was identified as a target for the authors of such primary studies. A notable exception are the 5 papers published in the ICEIS (International Conference on Enterprise Information Systems) conference and the 4 papers published in the CAiSE (International Conference on Advanced Information Systems Engineering) conference, which represent the 8\% and 6\% of the total, respectively. No papers were published in the BPM (International Conference on Business Process Management) conference. This may be explained by the reference role played by workflow patterns \cite{N.-Russell:2016aa}, rather than meta-models, in the evaluation of the expressiveness and adequacy of business process modelling languages. 

Concerning the research questions, which are the targets of this SLR, the numeric results and some comments are already contained in Section~\ref{sec:RRQs}. We report here some additional remarks that mainly highlight the overall findings and the limitations of current published research.

Focusing on \rqone, it is interesting to notice that half of the primary studies do not target any specific business process modelling language. This means that the description of what constitutes a business process is perceived as a topic of research \emph{per se}, and is not necessarily tight to a specific modelling language or approach. 
Also, most of the primary studies that
 focus on specific modelling languages target existing languages. This seems to indicate a reasonable maturity and level of satisfaction towards the available modelling languages, i.e., although there is still space of improvement in the field, a non-trivial amount of work has been carried out. 
By looking into the characteristics investigated in Table~\ref{table:Addtable} at page \pageref{table:Addtable} we can note that all
 the primary studies consider (at least) the process model dimension,
 with 38\% considering both. While the first result is perhaps not very surprising, we consider very positive the conspicuous presence of studies that incorporate also the execution dimension. Indeed, executions of processes are, in the BPM fields, regarded as first class citizens and not simply as mere instances of process models. As an example consider the importance of process executions (a.k.a
 event logs) in the field of Process Mining.
Another interesting result is the one that refers to the approaches taken towards business process modelling. As already said in Section~\ref{subsec:RQ1} 
most primary studies adhere to the traditional procedural and activity-centric based view on business processes with very few papers taking a only declarative, hybrid, or artefact-centric view (17\%, 5\% and 6\% of the primary studies, respectively).
Further interesting data concern the domain (in)dependency of meta-models. Indeed only about 5\% of the primary studies address domain dependent business process models. Thus, we can say that an effort to describe what constitutes a generic business process is well under way. Instead, investigations of what constitute a business process in a specific domain (e.g., an administrative procedure, a retail oriented business process, just to mention two popular domains) is way less clear and investigated.

The analysis of this research question provides us with an overall picture of what has been investigated in the literature.

Focusing on \rqtwo, a detailed analysis is already reported in Section~\ref{subsec:RQ2}. Summarising, the results shown in Table~\ref{elementsRQ2} at page \pageref{elementsRQ2}
indicate that the elements of the process control flow (\emph{activity} and \emph{sequence flow}) together with the \emph{data object}, the organisational dimension (\emph{actor}), and \emph{resource} are the most recurring both in terms of overall presence and decomposition into different elements. Instead \emph{goal} and \emph{value} aspects are poorly and, in case of \emph{value}, even not explicitly described both in terms of occurrences in primary studies meta-models and decomposition.
 This finding is certainly correlated to the fact that most business process modelling languages do not include values and/or goals in the graphical design of a business process model. Nonetheless, it is easy to observe that the situation does not change if we consider language-independent meta-models. This somehow clashes with most of the modern definitions of Business Process, which explicitly mention either the (added) value brought by the process execution\footnote{As an example, Johansson et al. \cite{Johansson:1993aa} say that a business process is ``''\emph{a set of linked activities that take an input and transform it to create an output. Ideally, the transformation that occurs in the process should add value to the input}''.} or the goal a process execution has to realise\footnote{See e.g., the definition taken from \cite{DBLP:books/daglib/0029914} and reported in Section~\ref{introduction} at page \pageref{def:process}.} 
Thus, while it seems to be ``extremely clear and well agreed that business processes realise a business goal", as recently highlighted in~\cite{AdamoAIIA2018}, it appears to be more difficult to leverage state-of-the-art business process meta-models to state exactly what this business goal (resp. value) is and which characteristics it detains~\cite{AdamoAIIA2018}.

If we focus on the elements appearing in at least 25\% of the studies, we can notice a high presence of elements related to the control flow w.r.t. other aspects of the business process. On the positive side, the elements related to the control flow that appear at least in 25\% of the studies\footnote{The control flow elements that appear in at least 25\% of the studies are \texttt{activity}, \texttt{atomic activity}, \texttt{compound activity}, \texttt{event}, \texttt{control flow}, \texttt{gateway}, \texttt{parallel gateway}, and \texttt{exclusive gateway}.} correspond to key elements of a business process control flow. On the negative side, in addition to the \texttt{goal} (\texttt{value}) aspect already discussed above, it is interesting to notice that while \texttt{actor} and \texttt{role} (as a \emph{resource}) are present in 25\%  of the studies, important elements such as \texttt{organization} and \texttt{artifact} are not. 
As already discussed at the end of Section~\ref{subsec:RQ2}, the element \texttt{resource} could also be considered a frequent element appearing in at least 25\% of meta-models. Nonetheless, it is important to notice that this would mean ignoring important differences that pertain the element itself such as having an agentive or a non-agentive characterisation, which we believe should be present when describing resources in the BPM context.
If we instead consider the elements appearing in at least 50\% of the studies, which could be considered a ``core'' set of elements of what constitutes a business process, we reduce to only \texttt{activity}. In our opinion this is a sign of a lack of a mature answer to the fundamental question of ``what constitutes a business process'' and an evidence of the fact that most works have mainly addressed business processes just looking at control flow related aspects, somehow neglecting a comprehensive investigation which takes into account all the characterising aspects of this notion. 

A clear issue comes out from the analysis of \rqtwo: some elements have an unclear and overloaded meaning, as in the case of \emph{resource} and \emph{event}.
As described in Subsection~\ref{subsec:RQ2}, some elements of the \emph{resource} macro-element, also appear in other categories such as \emph{actor} or \emph{data-object}. Let us consider the case of \texttt{information}. While 
intuitively we can see why this element is classified under the umbrella of both \emph{data-object} and \emph{resource}, we would argue that it is in both categories but for two different reasons: \texttt{information} ``is a'' \emph{data-object} that can ``play the role'' of a resource (and is not a resource per-se). A similar analysis could be done for the element \texttt{process participant} which lies across \emph{actor} and \emph{resource}. This kind of analysis can be fostered and motivated by results of systematic studies such as the one for answering \rqtwo. 
Similarly, the same term ``event'' can assume different meanings: (i) event $\grave{a}$-la BPMN, i.e., ``something that happens during the course of a process''~\cite{model2011notation}; (ii) event as \emph{state} describing pre- and and post-conditions~\cite{Mendling2008}, and (iii) a generic \emph{event} whose exact meaning should be further investigated with care. 

When focusing on the notion of \emph{state}, we observed a higher number of occurrences of precondition (17) with respect to the frequency of postcondition elements (9). Does it mean that preconditions for the executions of activities are, for the BPM community, more important than the effects of the activities themselves? If so, why? Related to this, one could also observe in Table~\ref{elementsRQ2}, a scarce presence of a notion of \emph{trigger}, i.e., of something that, differently from a state, can cause the activation of the activities. Is this larger occurrences of preconditions, only related to the specification of necessary conditions for the execution of an activity or is the notion of precondition sometimes used also for specifying a triggering power of this the precondition state? If this is the case, this should be investigated and clarified in an explicit manner as states, as commonly understood, do not have causal power. These are examples of further questions one should address to precisely understand what these core elements of business processes are.

If on the one hand, some element labels are overloaded, on the other hand, some elements have different labels but close semantics. While in some cases aggregating different labels with the same semantics is not difficult  (e.g., in the case of  ``gateway'' and ``decision diamond''), in other cases, when the semantics of the elements is close, though not identical, it is hard to decide whether the elements can be generalised under a unique term. An example of this is the case of \texttt{activity}, \texttt{transition}, and \texttt{function}. Indeed, although the semantics of these elements are quite close, differently from \texttt{activity}, \texttt{transition} and \texttt{function} have an explicit state transformation flavour.

The elicitation of the label overloading and the potential element generalisation, as well as the subsequent need to clarify what exactly the elements mean, are interesting aspects that emerge from this SLR and that, in our opinion, represent important challenges for the BPM community that could foster further research such as the one described in \citet{DBLP:conf/caise/AdamoFG20}, and also pave the way to the identification of clear, language-independent, reference meta-models.
Although the definition of a unique shared reference meta-model could be difficult, and even impossible, to achieve, the answer to this research question provides an initial building block towards the clarification of the elements that should be involved in the construction of reference meta-models that are however connected through well-investigated relationships and mappings. 
Shared reference meta-models would indeed be beneficial both for the process model understandability and for the process model design, by improving model understanding, interoperability among models, mapping between models designed by using different languages, as well as by improving the quality of the designed models.

The results of \rqthree, summarised in Table \ref{table:role}, show an interesting and articulated usage of business process meta-models. Even though the most popular usage of meta-models is somehow self referential (``extension of the meta-model with a new concept''), the number of other usages denote that a non-trivial amount of work has been carried out
 in the field, in particular for what concern the exploitation of meta-models to investigate aspects of specific business process modelling languages. A possible limitation here is the lack of foundational studies that address the fundamental question of what a business process is and what differentiates it from other kinds of processes. 

Indeed, this variety of meta-models, many of which extend other as many meta-models, suggests again that 
there is the need of one or more reference and agreed meta-models, that could provide a reference point for the community and that could be more useful than several different meta-models.  

The results of \rqfour, summarised in Table~\ref{table:evaluation}, are on the problematic side. Indeed only slightly more than 6\% of the primary studies show a sufficiently extensive evaluation phase with real case studies, and another 12\% show an evaluation of the adequacy of the meta-model using some sort of comparative analysis against reference bodies of knowledge (a reference foundational ontology, requirements, or alternative meta-models). This lack of coverage of the ``evaluation'' phase may be justified by many different factors: on the one hand evaluating the adequacy, or usefulness, of a generic meta-model in concrete domain-specific scenarios is a complex activity, especially when there are no standard reference scenarios for this activity; on the other hand meta-models are introduced for different purposes (see the answer to \rqthree) and different purposes may require different evaluation strategies. 
 This finding highlights a limitation of current research, and the BPM community should make an effort to face such a challenge in order to understand whether (and how) an evaluation of meta-models could be carried out. Nonetheless, the two typologies of evaluation present in the primary studies (evaluation with real case studies; comparative evaluation) plus the evaluation in terms of CMQF and semi-structured interviews provide a good starting point for this discussion.

Summing up, this analysis allowed us to identify important criticalities and challenges for the BPM community. Solid reference meta-models with a shared understanding of their elements and relationships and that can represent a reference point for extensions, new languages, as well as for defining and improving model quality are better than many different weak meta-models. A good evaluation strategy for evaluating these meta-models should also be defined and applied.

\subsection{Limitations of this study} 
\label{sub:limitations_of_this_study}

Part of the limitations of the study are related to its \emph{internal} validity and include: (i) biases in the selection of the papers; (ii) imprecisions introduced in the extraction of data from the selected works.
To mitigate these threats, we followed the guidelines reported in~\cite{Kitchenham07guidelinesfor,kitchenham2004procedures}. 
We applied the standard procedures reported in the guidelines for the correctness of the SLRs, such as the identification of the proper keywords to perform the data search, the selection of the appropriate sources and repositories for the field under investigation, the definition of clear inclusion and exclusion criteria, as well as of the quality assessment parameters.
Specifically, we relied on the main literature sources and libraries in the information system field for the extraction of the works related to business process models and meta-models. Moreover, we expanded the search by manually inspecting the two main reference conferences in the field of BPM. To further improve the reliability of the review, we put some effort in guaranteeing the reproducibility of the search by other researchers, although ranking algorithms used by the source libraries could be updated and provide different results. 

Another limitation is related to the \emph{external} validity of the study. Our search was limited to papers in English language, thus limiting the generalizability of the results. We indeed specified as part of the WoS query that papers should be in English language and we further applied an exclusion criteria to remove the non-English papers extracted from the other repositories. However, processing non-English papers would open the issues of which languages to include and of having the linguistic abilities to process non-English papers. Moreover, we expect that the literature in English is able to capture a significant picture of the scientific works carried out.

Concerning the \emph{construct} validity, the search method and the search query used for the automated paper extraction could have left out some relevant papers. In particular, we mainly focused on (i) manual search on two reference conferences, and (ii) database search using queries looking for explicit meta-models.
This may have left out papers that could be obtained by other search methods, in particular snowball search, or by query terms looking for implicit meta-models such as the ones derived from formal representations. This could be a limitation especially for \rqtwo. However, the first limitation is partly mitigated by exploiting two out of the three commonly used search methods for Systematic Literature Reviews~\cite{Brings2018}, i.e., manual and database search, and the second limitation is mitigated by the fact that whenever an explicit meta-model reflecting the corresponding formal specification is available, we are anyway able to indirectly capture the elements of the formal representation.

A further limitation hampering the \emph{conclusion} validity of this study is related to the potential inaccuracies due to the subjectivity of the analysis carried out. Indeed, (i) only one researcher selected the candidate primary studies;
 and  (ii) only one researcher worked on the data extraction. Nevertheless, both aspects have been mitigated by the fact that (i) another researcher checked the inclusion and the exclusion of the studies; and (ii) another researcher checked the data extraction, as suggested in~\cite{BRERETON2007571}.


\section{Conclusions}
\label{sec:conclusion}

This work provides the first systematic literature review of business process meta-models. This systematic literature review addressed research questions concerning (i) the kind of meta-models proposed in literature; (ii) the recurring constructs they contain; (iii) their purpose(s); and (iv) their evaluations. 

The analysis provided in this SLR shows that there is a reasonable body of work conducted in this specific area, even though the field does not appear to have reached full maturity. On the positive side, a reasonable number of high quality publications exist in literature, which present well described business process meta-models. These meta-models are almost equally targeting specific BPMLs or the notion of business process in general. Also, they cover both the model and execution aspects of business processes. Another positive aspect is related to the number of different reasons for introducing/exploiting these meta-models, which is an evidence of liveliness of the topic, and the reasonable presence of key control flow elements in the meta-models. Also, some good examples of how to evaluate meta-models are present in literature.   
On the negative side we can notice: a lack of meta-models for the ``new'' paradigms
towards business process modelling, namely, the declarative based and artefact centric approaches; a lack of presence of non control flow key aspects of business processes in meta-models; and a lack of evaluation of meta-models in literature.   
These results could open up an opportunity for new research efforts addressing these aspects.

The analysis provided in this SLR could be used as a starting point to define a rich framework for the description and classification of business process meta-models.
 Indeed, the characteristics identified in answering \rqone, \rqthree, and \rqfour provide an extensive set of ``tags'' which could be used to annotate meta-models, while analogous ``tags'' to describe the content could be defined starting from the answer to \rqtwo. This framework can be used not only for classification purposes but, in turn, also
 to retrieve meta-models with specific characteristics (e.g., with the purpose of identifying the most suitable meta-model or process modelling language for specific needs), to define a set of guidelines for well-defined meta-models, or to compare and analyse business process meta-models and languages further in the future.
Similarly, the analysis of the meta-model elements produced in answering \rqtwo could be used as a starting point for defining an ``emerging'' business process meta-model from data. To do that, an analysis of the relationships between these elements (or at least between the most recurring ones) should be produced, and this is part of a work we would like to start in the immediate future.  


\appendix
\section{Primary Studies' Publication Venues}
\label{app:sourcelist}
\begin{center}
\scriptsize
\begin{longtable}{p{10cm} p{1cm} }
	\toprule
	Journal & Paper \\ 
	\midrule
	Information and Software Technology$^{**}$ & \cite{DBLP:journals/infsof/ArevaloCRD16,
	DBLP:journals/infsof/StrembeckM11}\\
	International Journal of Business Process Integration and Management & 	\cite{DBLP:journals/ijbpim/AxenathKR07, DBLP:journals/ijbpim/RosemannRF08} \\
	Journal on Data Semantics$^{*}$ & \cite{DBLP:journals/jodsn/CherfiAC13} \\
 	Journal of Knowledge Management & \cite{DBLP:journals/jkm/PapavassiliouM03}\\
 	Information Systems$^{**}$ & \cite{DBLP:journals/is/DamaggioHV13} \\
	Procedia Computer Science$^{*}$ & \cite{martins2015business} \\
	Group Decision and Negotiation$^{*}$ & \cite{Farrell2006FormalisingW} \\
	Expert Systems with Applications$^{**}$ & \cite{DBLP:journals/eswa/MertensGP17} \\
	IEEE Access$^{**}$ & \cite{DBLP:journals/access/AmjadAABRN18} \\
	Decision and Support Systems$^{**}$ & \cite{DBLP:journals/dss/KwanB03} \\
	European Journal of Information Systems$^{**}$ & \cite{DBLP:journals/ejis/Rittgen06}\\
	Data and Knowledge Engineering$^{*}$ & \cite{10.1016/S0169-023X(01)00034-9}\\
	\toprule
	Conference \& Symposium & Paper \\ 
	\midrule
	International Conference on Business Informatics & \cite{DBLP:conf/wecwis/HeidariLBB13,DBLP:conf/wecwis/StratigakiNLA16} \\
	International Conference on Intelligent Systems Design and Applications & \cite{DBLP:conf/isda/HassenTG16} \\
	International Conference on Database and Expert Systems Applications$^{*}$ &	\cite{DBLP:conf/dexa/NicolaMPS10} \\
	International Conference on Conceptual Modeling$^{**}$ & \cite{DBLP:conf/er/RosaDHMG08} \\
	International Conference on Information Systems$^{**}$ & \cite{DBLP:conf/icis/HuaZS10} \\
	East European Conference on Advances in Databases and Information Systems & 	\cite{DBLP:conf/adbis/MomotkoS04} \\
	International Conference on Enterprise Information Systems & 	\cite{DBLP:conf/iceis/KorherrL07,DBLP:conf/iceis/BouneffaA13,DBLP:conf/iceis/AhnKK18,DBLP:conf/iceis/RamdoyalPDSLJ13,DBLP:conf/iceis/CombemaleCGTV07} \\
	International Conference on Advanced Information Systems Engineering$^{**}$ & \cite{DBLP:conf/caise/SoderstromAJPW02,DBLP:conf/caise/RuizCEFP14,DBLP:conf/caise/RussellAHE05, DBLP:conf/caise/WeigandJABEI06} \\
	International Conference on Enterprise Systems  & \cite{DBLP:conf/es/SprovieriV15, DBLP:conf/es/FanesiCH15} \\
	Hawaii International Conference on System Sciences & \cite{DBLP:conf/hicss/WeissW11a,DBLP:conf/hicss/FriedenstabJMM12} \\
	International Conference on Information Integration and Web-based Applications and Services & \cite{DBLP:conf/iiwas/YahyaBMB16}\\
	International Conference on Computer Theory and Applications & \cite{Mahdi2012} \\
	International Conference on Commerce and Enterprise Computing$^{*}$ & \cite{DBLP:conf/wecwis/ReddingDHI09}\\
	International Conference on Knowledge-Based and Intelligent Information {\&} Engineering Systems$^{*}$ & \cite{DBLP:conf/kes/OualiMB16}\\
	International Conference on World Wide Web$^{**}$ & \cite{DBLP:conf/www/HallerOK06}\\
	International Conference on Bioinformatics and Biomedicine & \cite{DBLP:conf/bibm/BraunSBE14}\\
	International Business Information Management Association Conference$^{*}$ & \cite{Thabet2015}\\
	International Conference on Information Technology Interfaces & \cite{Groznic2002}\\
	International Conference on Web Services$^{**}$ & \cite{Albert2005}\\
	Working Conference on The Practice of Enterprise Modeling & \cite{DBLP:conf/ifip8-1/GaoK10} \\
	International Enterprise Distributed Object Computing Conference$^{*}$ & 	\cite{DBLP:conf/edoc/BruningG11} \\
	Ibero-American Conference on Software Engineering & \cite{Stroppi2011}\\
	East European Conference on Advances in Databases and Information Systems$^{*}$ & \cite{DBLP:conf/adbis/EderG02}\\	
	Multikonferenz Wirtschaftsinformatik & \cite{krumeich2016modeling}\\
	Internationale Tagung Wirtschaftsinformatik & \cite{DBLP:conf/wi/DorndorferS17} \\
	ACM Symposium on Applied Computing$^{*}$ & \cite{DBLP:conf/sac/ListK06,DBLP:conf/sac/SantosAG10}\\
	International Symposium on Business Modeling and Software Design & 	\cite{hassen2017extending} \\
	European Semantic Web Symposium on The Semantic Web: Research and Applications$^{**}$ & \cite{DBLP:conf/esws/GasevicD04}\\
	\toprule
	Workshop  & Paper \\ 
	\midrule	
	International Workshop on Personalization and Context-Awareness in Cloud and Service Computing & 	\cite{DBLP:conf/wise/KunchalaYY14}\\
	Workshop on Enterprise and Organizational Modeling and Simulation & 	\cite{DBLP:conf/caise/HeidariLK10} \\
	Workshop XML for Business Process Management & \cite{thom2005improving} \\
	International Workshop on Business Process Modeling Notation & 	\cite{DBLP:conf/bpmn/Natschlager11} \\
	International Workshop on Enterprise {\&} Organizational Modeling and Simulation & \cite{Bouchbout2010} \\
	International Workshop on Web Information and Data Management & \cite{DBLP:conf/widm/BernardiCLM12}\\
	International Workshop on Best Practices of UML & \cite{DBLP:conf/er/ListK05}\\
	International Workshop on Business Process Modeling, Development and Support & \cite{DBLP:conf/caise/BessaiN09}\\
	International Workshop on Models and Model-driven Methods for Enterprise Computing & \cite{DBLP:conf/edoc/MilanovicGW08}\\
	International Workshop on Object-Role Modeling & \cite{DBLP:conf/otm/GoedertierV07}\\
	International Workshop on Adequacy of Modelling Methods & \cite{DBLP:conf/modellierung/JannaberKRTDB16} \\
	Workshop on Interoperability of Heterogeneous Information Systems & \cite{DBLP:conf/ihis/HamriBB05}\\
	\bottomrule
\end{longtable}
\end{center}

The venues marked with $^{**}$ are classified as Quartile 1 (Q1) or A/A$^*$ according to the Scopus journal ranking 2017 and the CORE conference ranking 2017, respectively.   
The venues marked with $^{*}$ are classified as Quartile 2 (Q2) or B according to the Scopus journal ranking 2017 and the CORE conference ranking 2017, respectively.



\begin{thebibliography}{104}
\providecommand{\natexlab}[1]{#1}
\providecommand{\url}[1]{\texttt{#1}}
\expandafter\ifx\csname urlstyle\endcsname\relax
  \providecommand{\doi}[1]{doi: #1}\else
  \providecommand{\doi}{doi: \begingroup \urlstyle{rm}\Url}\fi

\bibitem[Adamo et~al.(2017)Adamo, Borgo, {Di Francescomarino}, Ghidini,
  Guarino, and Sanfilippo]{Adamo:2017:AIIA}
Greta Adamo, Stefano Borgo, Chiara {Di Francescomarino}, Chiara Ghidini, Nicola
  Guarino, and Emilio~M. Sanfilippo.
\newblock Business processes and their participants: An ontological
  perspective.
\newblock In \emph{Proceedings of the 16th International Conference of the
  Italian Association for Artificial Intelligence (AI*IA 2017)}, volume 10640
  of \emph{\nbtex{Lecture Notes in Computer Science}}, pages 215--228. Springer
  International Publishing, 2017.
\newblock ISBN 978-3-319-70169-1.

\bibitem[Adamo et~al.(2018)Adamo, Borgo, {Di Francescomarino}, Ghidini, and
  Guarino]{AdamoAIIA2018}
Greta Adamo, Stefano Borgo, Chiara {Di Francescomarino}, Chiara Ghidini, and
  Nicola Guarino.
\newblock On the notion of goal in business process models.
\newblock In \emph{AI*IA 2018 - Advances in Artificial Intelligence - XVIIth
  International Conference of the Italian Association for Artificial
  Intelligence, Trento, Italy, November 20-23, 2018, Proceedings}, volume 11298
  of \emph{Lecture Notes in Computer Science}, pages 139--151. Springer, 2018.
\newblock ISBN 978-3-030-03839-7.
\newblock \doi{10.1007/978-3-030-03840-3}.

\bibitem[Adamo et~al.(2019)Adamo, Ghidini, and {Di
  Francescomarino}]{adamo2019whats}
Greta Adamo, Chiara Ghidini, and Chiara {Di Francescomarino}.
\newblock {What's My Process Model Composed of? A Systematic Literature Review
  of Meta-Models in BPM}.
\newblock ArXiv abs/1910.05564, 2019.
\newblock Submitted for publication.

\bibitem[Adamo et~al.(2020)Adamo, {Di Francescomarino}, and
  Ghidini]{DBLP:conf/caise/AdamoFG20}
Greta Adamo, Chiara {Di Francescomarino}, and Chiara Ghidini.
\newblock Digging into business process meta-models: {A} first ontological
  analysis.
\newblock In Schahram Dustdar, Eric Yu, Camille Salinesi, Dominique Rieu, and
  Vik Pant, editors, \emph{Advanced Information Systems Engineering - 32nd
  International Conference, CAiSE 2020, Grenoble, France, June 8-12, 2020,
  Proceedings}, volume 12127 of \emph{Lecture Notes in Computer Science}, pages
  384--400. Springer, 2020.
\newblock \doi{10.1007/978-3-030-49435-3\_24}.

\bibitem[Aguilar-Savén(2004)]{AGUILARSAVEN2004129}
Ruth~Sara Aguilar-Savén.
\newblock Business process modelling: Review and framework.
\newblock \emph{International Journal of Production Economics}, 90\penalty0
  (2):\penalty0 129 -- 149, 2004.
\newblock ISSN 0925-5273.
\newblock \doi{https://doi.org/10.1016/S0925-5273(03)00102-6}.

\bibitem[Ahn et~al.(2018)Ahn, Kim, and Kim]{DBLP:conf/iceis/AhnKK18}
Hyun Ahn, Kyoungsook Kim, and Kwanghoon~Pio Kim.
\newblock Generating process entity hierarchies from {XPDL} process models.
\newblock In \emph{Proceedings of the 20th International Conference on
  Enterprise Information Systems, {ICEIS} 2018, Funchal, Madeira, Portugal,
  March 21-24, 2018, Volume 2}, pages 690--695. SciTePress, 2018.

\bibitem[{Albert} et~al.(2005){Albert}, {Henocque}, and {Kleiner}]{Albert2005}
P.~{Albert}, L.~{Henocque}, and M.~{Kleiner}.
\newblock Configuration based workflow composition.
\newblock In \emph{IEEE International Conference on Web Services (ICWS'05)},
  volume~1, pages 285--292, 2005.

\bibitem[Amjad et~al.(2018)Amjad, Azam, Anwar, Butt, Rashid, and
  Naeem]{DBLP:journals/access/AmjadAABRN18}
Anam Amjad, Farooque Azam, Muhammad~Waseem Anwar, Wasi~Haider Butt, Muhammad
  Rashid, and Aamir Naeem.
\newblock {UMLPACE} for modeling and verification of complex business
  requirements in event-driven process chain {(EPC)}.
\newblock \emph{{IEEE} Access}, 6:\penalty0 76198--76216, 2018.

\bibitem[Ar{\'{e}}valo et~al.(2016)Ar{\'{e}}valo, Escalona, Ramos, and
  Dom{\'{\i}}nguez{-}Mu{\~{n}}oz]{DBLP:journals/infsof/ArevaloCRD16}
Carlos Ar{\'{e}}valo, Mar{\'{\i}}a~Jos{\'{e}} Escalona, Isabel Ramos, and
  M.~Dom{\'{\i}}nguez{-}Mu{\~{n}}oz.
\newblock A metamodel to integrate business processes time perspective in
  {BPMN} 2.0.
\newblock \emph{Information {\&} Software Technology}, 77:\penalty0 17--33,
  2016.

\bibitem[Axenath et~al.(2007)Axenath, Kindler, and
  Rubin]{DBLP:journals/ijbpim/AxenathKR07}
Bj{\"{o}}rn Axenath, Ekkart Kindler, and Vladimir~A. Rubin.
\newblock {AMFIBIA:} a meta-model for integrating business process modelling
  aspects.
\newblock \emph{{International Journal of Business Process Integration and
  Management}}, 2\penalty0 (2):\penalty0 120--131, 2007.

\bibitem[{Ben Hassen} et~al.(2016){Ben Hassen}, Turki, and
  Gargouri]{DBLP:conf/isda/HassenTG16}
Mariam {Ben Hassen}, Mohamed Turki, and Fa{\"{\i}}ez Gargouri.
\newblock A proposal to model knowledge dimension in sensitive business
  processes.
\newblock In \emph{Intelligent Systems Design and Applications - 16th
  International Conference on Intelligent Systems Design and Applications
  {(ISDA} 2016) held in Porto, Portugal, December 16-18, 2016}, volume 557 of
  \emph{Advances in Intelligent Systems and Computing}, pages 1015--1030.
  Springer, 2016.
\newblock ISBN 978-3-319-53479-4.

\bibitem[{Ben Hassen} et~al.(2017){Ben Hassen}, Turki, and
  Gargouri]{hassen2017extending}
Mariam {Ben Hassen}, Mohamed Turki, and Fa{\"{\i}}ez Gargouri.
\newblock Extending bpmn 2.0 with the knowledge dimension.
\newblock In \emph{Proceedings of the Seventh International Symposium on
  Business Modeling and Software Design}, pages 111--124. SciTePress, 2017.
\newblock ISBN 978-989-758-238-7.

\bibitem[Bernardi et~al.(2012)Bernardi, Cimitile, {Di Lucca}, and
  Maggi]{DBLP:conf/widm/BernardiCLM12}
Mario~Luca Bernardi, Marta Cimitile, Giuseppe~A. {Di Lucca}, and Fabrizio~Maria
  Maggi.
\newblock {M3D:} a tool for the model driven development of web applications.
\newblock In George H.~L. Fletcher and Prasenjit Mitra, editors,
  \emph{Proceedings of the Twelfth International Workshop on Web Information
  and Data Management, {WIDM} 2012, Maui, HI, USA, November 02, 2012}, pages
  73--80. {ACM}, 2012.

\bibitem[Bessai and Nurcan(2009)]{DBLP:conf/caise/BessaiN09}
Kahina Bessai and Selmin Nurcan.
\newblock Actor-driven approach for business process. how to take into account
  the work environment?
\newblock In Terry~A. Halpin, John Krogstie, Selmin Nurcan, Erik Proper, Rainer
  Schmidt, Pnina Soffer, and Roland Ukor, editors, \emph{Enterprise,
  Business-Process and Information Systems Modeling, 10th International
  Workshop, {BPMDS} 2009, and 14th International Conference, {EMMSAD} 2009,
  held at CAiSE 2009, Amsterdam, The Netherlands, June 8-9, 2009. Proceedings},
  volume~29 of \emph{Lecture Notes in Business Information Processing}, pages
  187--196. Springer, 2009.

\bibitem[Bhattacharya et~al.(2009)Bhattacharya, Hull, and
  Su]{Bhattacharya09adata-centric}
Kamal Bhattacharya, Richard Hull, and Jianwen Su.
\newblock A data-centric design methodology for business processes.
\newblock In \emph{Handbook of Research on Business Process Modeling},
  chapter~23, pages 503--531. Igi-global, 2009.

\bibitem[Bouchbout et~al.(2010)Bouchbout, Akoka, and Alimazighi]{Bouchbout2010}
Khoutir Bouchbout, Jacky Akoka, and Zaia Alimazighi.
\newblock Proposition of a generic metamodel for interorganizational business
  processes.
\newblock In \emph{Proceedings of the 6th International Workshop on Enterprise
  {\&} Organizational Modeling and Simulation}, EOMAS ’10, page 42–56,
  Aachen, DEU, 2010. CEUR-WS.org.
\newblock ISBN 9781450304634.

\bibitem[Bouneffa and Ahmad(2013)]{DBLP:conf/iceis/BouneffaA13}
Mourad Bouneffa and Adeel Ahmad.
\newblock Change management of bpm-based software applications.
\newblock In \emph{{ICEIS} 2013 - Proceedings of the 15th International
  Conference on Enterprise Information Systems, Volume 2, Angers, France, 4-7
  July, 2013}, pages 37--45. SciTePress, 2013.
\newblock ISBN 978-989-8565-60-0.

\bibitem[Braun et~al.(2014)Braun, Schlieter, Burwitz, and
  Esswein]{DBLP:conf/bibm/BraunSBE14}
Richard Braun, Hannes Schlieter, Martin Burwitz, and Werner Esswein.
\newblock {BPMN4CP:} design and implementation of a {BPMN} extension for
  clinical pathways.
\newblock In Huiru~Jane Zheng, Werner Dubitzky, Xiaohua Hu, Jin{-}Kao Hao,
  Daniel~P. Berrar, Kwang{-}Hyun Cho, Yadong Wang, and David~R. Gilbert,
  editors, \emph{2014 {IEEE} International Conference on Bioinformatics and
  Biomedicine, {BIBM} 2014, Belfast, United Kingdom, November 2-5, 2014}, pages
  9--16. {IEEE} Computer Society, 2014.

\bibitem[Brereton et~al.(2007)Brereton, Kitchenham, Budgen, Turner, and
  Khalil]{BRERETON2007571}
Pearl Brereton, Barbara~A. Kitchenham, David Budgen, Mark Turner, and Mohamed
  Khalil.
\newblock Lessons from applying the systematic literature review process within
  the software engineering domain.
\newblock \emph{Journal of Systems and Software}, 80\penalty0 (4):\penalty0 571
  -- 583, 2007.
\newblock ISSN 0164-1212.
\newblock \doi{https://doi.org/10.1016/j.jss.2006.07.009}.
\newblock Software Performance.

\bibitem[Brings et~al.(2018)Brings, Daun, Kempe, and Weyer]{Brings2018}
Jennifer Brings, Marian Daun, Markus Kempe, and Thorsten Weyer.
\newblock On different search methods for systematic literature reviews and
  maps: Experiences from a literature search on validation and verification of
  emergent behavior.
\newblock In \emph{Proceedings of the 22nd International Conference on
  Evaluation and Assessment in Software Engineering 2018}, EASE'18, page
  35–45, New York, NY, USA, 2018. Association for Computing Machinery.
\newblock ISBN 9781450364034.
\newblock \doi{10.1145/3210459.3210463}.

\bibitem[Br{\"{u}}ning and Gogolla(2011)]{DBLP:conf/edoc/BruningG11}
Jens Br{\"{u}}ning and Martin Gogolla.
\newblock {UML} metamodel-based workflow modeling and execution.
\newblock In \emph{Proceedings of the 15th {IEEE} International Enterprise
  Distributed Object Computing Conference, {EDOC} 2011, Helsinki, Finland,
  August 29 - September 2, 2011}, pages 97--106. {IEEE} Computer Society, 2011.
\newblock ISBN 978-1-4577-0362-1.

\bibitem[Cherfi et~al.(2013)Cherfi, Ayad, and
  Comyn{-}Wattiau]{DBLP:journals/jodsn/CherfiAC13}
Samira~Si{-}Said Cherfi, Sarah Ayad, and Isabelle Comyn{-}Wattiau.
\newblock Improving business process model quality using domain ontologies.
\newblock \emph{J. Data Semantics}, 2\penalty0 (2-3):\penalty0 75--87, 2013.

\bibitem[Combemale et~al.(2007)Combemale, Cr{\'{e}}gut, Garoche, Thirioux, and
  Vernadat]{DBLP:conf/iceis/CombemaleCGTV07}
Beno{\^{\i}}t Combemale, Xavier Cr{\'{e}}gut, Pierre{-}Lo{\"{\i}}c Garoche,
  Xavier Thirioux, and Fran{\c{c}}ois Vernadat.
\newblock A property-driven approach to formal verification of process models.
\newblock In Joaquim Filipe, Jos{\'{e}} Cordeiro, and Jorge~S. Cardoso,
  editors, \emph{Enterprise Information Systems, 9th International Conference,
  {ICEIS} 2007, Funchal, Madeira, Portugal, June 12-16, 2007, Revised Selected
  Papers}, volume~12 of \emph{Lecture Notes in Business Information
  Processing}, pages 286--300. Springer, 2007.

\bibitem[da~Silva(2015)]{Silva2015ModeldrivenEA}
Alberto~Rodrigues da~Silva.
\newblock Model-driven engineering: A survey supported by the unified
  conceptual model.
\newblock \emph{Computer Languages, Systems {\&} Structures}, 43:\penalty0
  139--155, 2015.

\bibitem[Damaggio et~al.(2013)Damaggio, Hull, and
  Vacul{\'{\i}}n]{DBLP:journals/is/DamaggioHV13}
Elio Damaggio, Richard Hull, and Roman Vacul{\'{\i}}n.
\newblock On the equivalence of incremental and fixpoint semantics for business
  artifacts with guard-stage-milestone lifecycles.
\newblock \emph{Inf. Syst.}, 38\penalty0 (4):\penalty0 561--584, 2013.

\bibitem[Davis(2001)]{EPC-Davis}
Rob Davis.
\newblock \emph{Business Process Modelling with ARIS - A Practical Guide}.
\newblock Springer, 2001.

\bibitem[{De Giacomo} et~al.(2015){De Giacomo}, Dumas, Maggi, and
  Montali]{DBLP:conf/caise/GiacomoDMM15}
Giuseppe {De Giacomo}, Marlon Dumas, Fabrizio~Maria Maggi, and Marco Montali.
\newblock Declarative process modeling in {BPMN}.
\newblock In \emph{Advanced Information Systems Engineering - 27th
  International Conference, CAiSE 2015, Stockholm, Sweden, June 8-12, 2015,
  Proceedings}, volume 9097 of \emph{Lecture Notes in Computer Science}, pages
  84--100. Springer, 2015.
\newblock ISBN 978-3-319-19068-6.

\bibitem[{De Nicola} et~al.(2007){De Nicola}, Lezoche, and
  Missikoff]{DBLP:conf/iicai/NicolaLM07}
Antonio {De Nicola}, Mario Lezoche, and Michele Missikoff.
\newblock An ontological approach to business process modeling.
\newblock In \emph{Proceedings of the 3rd Indian International Conference on
  Artificial Intelligence, Pune, India, December 17-19, 2007}, pages
  1794--1813. {IICAI}, 2007.
\newblock ISBN 978-0-9727412-2-4.

\bibitem[{De Nicola} et~al.(2010){De Nicola}, Missikoff, Proietti, and
  Smith]{DBLP:conf/dexa/NicolaMPS10}
Antonio {De Nicola}, Michele Missikoff, Maurizio Proietti, and Fabrizio Smith.
\newblock An open platform for business process modeling and verification.
\newblock In \emph{Database and Expert Systems Applications, 21st International
  Conference, {DEXA} 2010, Bilbao, Spain, August 30 - September 3, 2010,
  Proceedings, Part {I}}, volume 6261 of \emph{Lecture Notes in Computer
  Science}, pages 76--90. Springer, 2010.
\newblock ISBN 978-3-642-15363-1.

\bibitem[D{\"{o}}rndorfer and Seel(2017)]{DBLP:conf/wi/DorndorferS17}
Julian D{\"{o}}rndorfer and Christian Seel.
\newblock A meta model based extension of {BPMN} 2.0 for mobile context
  sensitive business processes and applications.
\newblock In \emph{Towards Thought Leadership in Digital Transformation: 13.
  Internationale Tagung Wirtschaftsinformatik, {WI} 2017, St.Gallen,
  Switzerland, February 12-15, 2017.}, 2017.

\bibitem[Eder and Gruber(2002)]{DBLP:conf/adbis/EderG02}
Johann Eder and Wolfgang Gruber.
\newblock A meta model for structured workflows supporting workflow
  transformations.
\newblock In Yannis Manolopoulos and Pavol N{\'{a}}vrat, editors,
  \emph{Advances in Databases and Information Systems, 6th East European
  Conference, {ADBIS} 2002, Bratislava, Slovakia, September 8-11, 2002,
  Proceedings}, volume 2435 of \emph{Lecture Notes in Computer Science}, pages
  326--339. Springer, 2002.

\bibitem[Elgammal et~al.(2016)Elgammal, T{\"{u}}retken, van~den Heuvel, and
  Papazoglou]{DBLP:journals/sosym/ElgammalTHP16}
Amal Elgammal, Oktay T{\"{u}}retken, Willem{-}Jan van~den Heuvel, and Mike~P.
  Papazoglou.
\newblock Formalizing and appling compliance patterns for business process
  compliance.
\newblock \emph{Software and Systems Modeling}, 15\penalty0 (1):\penalty0
  119--146, 2016.
\newblock \doi{10.1007/s10270-014-0395-3}.

\bibitem[Fanesi et~al.(2015)Fanesi, Cacciagrano, and
  Hinkelmann]{DBLP:conf/es/FanesiCH15}
Diego Fanesi, Diletta~Romana Cacciagrano, and Knut Hinkelmann.
\newblock Semantic business process representation to enhance the degree of
  {BPM} mechanization - an ontology.
\newblock In \emph{International Conference on Enterprise Systems, {ES} 2015,
  Basel, Switzerland, October 14-15, 2015}, pages 21--32. {IEEE}, 2015.

\bibitem[Farrell et~al.(2007)Farrell, Sergot, and
  Bartolini]{Farrell2006FormalisingW}
Andrew D.~H. Farrell, Marek~J. Sergot, and Claudio Bartolini.
\newblock {Formalising Workflow : A CCS-inspired Characterisation of the YAWL
  Workflow Patterns}.
\newblock \emph{Group Decision and Negotiation}, 16\penalty0 (3):\penalty0
  213--254, May 2007.

\bibitem[Friedenstab et~al.(2012)Friedenstab, Janiesch, Matzner, and
  M{\"{u}}ller]{DBLP:conf/hicss/FriedenstabJMM12}
Jan{-}Philipp Friedenstab, Christian Janiesch, Martin Matzner, and Oliver
  M{\"{u}}ller.
\newblock Extending {BPMN} for business activity monitoring.
\newblock In \emph{45th Hawaii International International Conference on
  Systems Science {(HICSS-45} 2012), Proceedings, 4-7 January 2012, Grand
  Wailea, Maui, HI, {USA}}, pages 4158--4167. {IEEE} Computer Society, 2012.
\newblock \doi{10.1109/HICSS.2012.276}.

\bibitem[Gao and Krogstie(2010)]{DBLP:conf/ifip8-1/GaoK10}
Shang Gao and John Krogstie.
\newblock A repository architecture for business process characterizing models.
\newblock In Patrick van Bommel, Stijn Hoppenbrouwers, Sietse Overbeek, Erik
  Proper, and Joseph Barjis, editors, \emph{The Practice of Enterprise Modeling
  - Third {IFIP} {WG} 8.1 Working Conference, PoEM 2010, Delft, The
  Netherlands, November 9-10, 2010. Proceedings}, volume~68 of \emph{Lecture
  Notes in Business Information Processing}, pages 162--176. Springer, 2010.

\bibitem[Ga{\v{s}evi{\'{c}}} and
  Deved{\v{z}}i{\'{c}}(2004)]{DBLP:conf/esws/GasevicD04}
Dragan Ga{\v{s}evi{\'{c}}} and Vladan Deved{\v{z}}i{\'{c}}.
\newblock Reusing petri nets through the semantic web.
\newblock In Christoph Bussler, John Davies, Dieter Fensel, and Rudi Studer,
  editors, \emph{The Semantic Web: Research and Applications, First European
  Semantic Web Symposium, {ESWS} 2004, Heraklion, Crete, Greece, May 10-12,
  2004, Proceedings}, volume 3053 of \emph{Lecture Notes in Computer Science},
  pages 284--298. Springer, 2004.

\bibitem[Ghidini et~al.(2012)Ghidini, Rospocher, and
  Serafini]{Ghidini12serafini:modeling}
Chiara Ghidini, Marco Rospocher, and Luciano Serafini.
\newblock Serafini: Modeling in a wiki with moki: Reference architecture.
\newblock \emph{International Journal On Advances in Life Sciences}, 4\penalty0
  (3):\penalty0 111--124, 2012.

\bibitem[Goedertier and Vanthienen(2007)]{DBLP:conf/otm/GoedertierV07}
Stijn Goedertier and Jan Vanthienen.
\newblock Declarative process modeling with business vocabulary and business
  rules.
\newblock In Robert Meersman, Zahir Tari, and Pilar Herrero, editors, \emph{On
  the Move to Meaningful Internet Systems 2007: {OTM} 2007 Workshops, {OTM}
  Confederated International Workshops and Posters, AWeSOMe, CAMS, {OTM}
  Academy Doctoral Consortium, MONET, OnToContent, ORM, PerSys, PPN, RDDS,
  SSWS, and {SWWS} 2007, Vilamoura, Portugal, November 25-30, 2007,
  Proceedings, Part {I}}, volume 4805 of \emph{Lecture Notes in Computer
  Science}, pages 603--612. Springer, 2007.

\bibitem[Gonz{\'a}lez and Cabot(2014)]{Gonzlez2014FormalVO}
Carlos~A. Gonz{\'a}lez and Jordi Cabot.
\newblock Formal verification of static software models in mde: A systematic
  review.
\newblock \emph{Information {\&} Software Technology}, 56:\penalty0 821--838,
  2014.

\bibitem[{Groznik} and {Kova\v{c}i\v{c}}(2002)]{Groznic2002}
A.~{Groznik} and A.~{Kova\v{c}i\v{c}}.
\newblock Business renovation: from business process modelling to information
  system modelling.
\newblock In \emph{ITI 2002. Proceedings of the 24th International Conference
  on Information Technology Interfaces (IEEE Cat. No.02EX534)}, pages 405--409
  vol.1, 2002.

\bibitem[Guizzardi(2005)]{32d47c9b496748ff911fb88726992462}
Giancarlo Guizzardi.
\newblock \emph{Ontological foundations for structural conceptual models}.
\newblock PhD thesis, University of Twente, October 2005.

\bibitem[Guizzardi and Wagner(2005)]{Guizzardi-Wagner-UFO}
Giancarlo Guizzardi and Gerd Wagner.
\newblock A unified foundational ontology and some applications of it in
  business modeling.
\newblock In Peter~F. Green and Michael Rosemann, editors, \emph{Business
  Systems Analysis with Ontologies}, pages 345--367. IGI Global, 2005.

\bibitem[Haller et~al.(2006)Haller, Oren, and
  Kotinurmi]{DBLP:conf/www/HallerOK06}
Armin Haller, Eyal Oren, and Paavo Kotinurmi.
\newblock An ontology for internal and external business processes.
\newblock In Les Carr, David {De Roure}, Arun Iyengar, Carole~A. Goble, and
  Michael Dahlin, editors, \emph{Proceedings of the 15th international
  conference on World Wide Web, {WWW} 2006, Edinburgh, Scotland, UK, May 23-26,
  2006}, pages 1055--1056. {ACM}, 2006.

\bibitem[Hamri et~al.(2005)Hamri, Boufaida, and
  Boudjlida]{DBLP:conf/ihis/HamriBB05}
Salah Hamri, Mahmoud Boufaida, and Nacer Boudjlida.
\newblock An architecture for the interoperability of workflow models.
\newblock In Axel Hahn, Sven Abels, and Liane Haak, editors, \emph{Proceedings
  of the first international {ACM} workshop on Interoperability of
  Heterogeneous Information Systems (IHIS'05), {CIKM} Conference, Bremen,
  Germany, November 4, 2005}, pages 31--38. {ACM}, 2005.

\bibitem[Heidari et~al.(2011)Heidari, Loucopoulos, and
  Kedad]{DBLP:conf/caise/HeidariLK10}
Farideh Heidari, Pericles Loucopoulos, and Zoubida Kedad.
\newblock A quality-oriented business process meta-model.
\newblock In \emph{Enterprise and Organizational Modeling and Simulation - 7th
  International Workshop, {EOMAS} 2011, held at CAiSE 2011, London, UK, June
  20-21, 2011. Selected Papers}, volume~88 of \emph{Lecture Notes in Business
  Information Processing}, pages 85--99. Springer, 2011.

\bibitem[Heidari et~al.(2013)Heidari, Loucopoulos, Brazier, and
  Barjis]{DBLP:conf/wecwis/HeidariLBB13}
Farideh Heidari, Pericles Loucopoulos, Frances M.~T. Brazier, and Joseph
  Barjis.
\newblock A meta-meta-model for seven business process modeling languages.
\newblock In \emph{{IEEE} 15th Conference on Business Informatics, {CBI} 2013,
  Vienna, Austria, July 15-18, 2013}, pages 216--221. {IEEE} Computer Society,
  2013.
\newblock ISBN 978-0-7695-5072-5.

\bibitem[Hua et~al.(2010)Hua, Zhao, and Storey]{DBLP:conf/icis/HuaZS10}
Zhimin Hua, J.~Leon Zhao, and Veda~C. Storey.
\newblock Exploring a domain ontology based approach to business process
  design.
\newblock In \emph{Proceedings of the International Conference on Information
  Systems, {ICIS} 2010, Saint Louis, Missouri, USA, December 12-15, 2010}.
  Association for Information Systems, 2010.
\newblock ISBN 978-0-615-41898-8.

\bibitem[Jannaber et~al.(2016)Jannaber, Karhof, Riehle, Thomas, Delfmann, and
  Becker]{DBLP:conf/modellierung/JannaberKRTDB16}
Sven Jannaber, Arne Karhof, Dennis~M. Riehle, Oliver Thomas, Patrick Delfmann,
  and J{\"{o}}rg Becker.
\newblock Invigorating event-driven process chains - towards an integrated meta
  model for {EPC} standardization.
\newblock In Stefanie Betz and Ulrich Reimer, editors, \emph{Modellierung 2016,
  Workshopband}, volume {P-255} of \emph{{LNI}}, pages 13--22. {GI}, 2016.

\bibitem[Johansson et~al.(1993)Johansson, McHugh, Pendlebury, and
  Wheeler]{Johansson:1993aa}
Henry~J. Johansson, Patrick McHugh, A.~John Pendlebury, and William~A. Wheeler.
\newblock \emph{Business Process Reengineering: Breakpoint Strategies for
  Market Dominance}.
\newblock John Wiley \& Sons, 1993.

\bibitem[Kitchenham and Charters(2007)]{Kitchenham07guidelinesfor}
B.~Kitchenham and S~Charters.
\newblock Guidelines for performing systematic literature reviews in software
  engineering.
\newblock Technical Report EBSE 2007-001, Keele University and Durham
  University Joint Report, 2007.

\bibitem[Kitchenham(2004)]{kitchenham2004procedures}
Barbara Kitchenham.
\newblock Procedures for performing systematic reviews.
\newblock \emph{Keele, UK, Keele University}, 33\penalty0 (2004):\penalty0
  1--26, 2004.

\bibitem[Korherr and List(2007)]{DBLP:conf/iceis/KorherrL07}
Birgit Korherr and Beate List.
\newblock Extending the {EPC} and the {BPMN} with business process goals and
  performance measures.
\newblock In \emph{{ICEIS} 2007 - Proceedings of the Ninth International
  Conference on Enterprise Information Systems, Volume EIS, Funchal, Madeira,
  Portugal, June 12-16, 2007}, pages 287--294, 2007.
\newblock ISBN 978-972-8865-90-0.

\bibitem[Krumeich et~al.(2016)Krumeich, Zapp, Mayer, Werth, and
  Loos]{krumeich2016modeling}
Julian Krumeich, Manuel Zapp, Dirk Mayer, Dirk Werth, and Peter Loos.
\newblock Modeling complex event patterns in epc-models and transforming them
  into an executable event pattern language.
\newblock \emph{Multikonferenz Wirtschaftsinformatik (MKWI)}, pages 81--92,
  2016.

\bibitem[Kunchala et~al.(2014)Kunchala, Yu, and
  Yongchareon]{DBLP:conf/wise/KunchalaYY14}
Jyothi Kunchala, Jian Yu, and Sira Yongchareon.
\newblock A survey on approaches to modeling artifact-centric business
  processes.
\newblock In \emph{Web Information Systems Engineering - {WISE} 2014 Workshops
  - 15th International Workshops {IWCSN} 2014, Org2 2014, {PCS} 2014, and
  {QUAT} 2014, Thessaloniki, Greece, October 12-14, 2014, Revised Selected
  Papers}, volume 9051 of \emph{Lecture Notes in Computer Science}, pages
  117--132. Springer, 2014.
\newblock ISBN 978-3-319-20369-0.

\bibitem[Kwan and Balasubramanian(2003)]{DBLP:journals/dss/KwanB03}
M.~Millie Kwan and P.~R. Balasubramanian.
\newblock Knowledgescope: managing knowledge in context.
\newblock \emph{Decis. Support Syst.}, 35\penalty0 (4):\penalty0 467--486,
  2003.

\bibitem[{La Rosa} et~al.(2008){La Rosa}, Dumas, ter Hofstede, Mendling, and
  Gottschalk]{DBLP:conf/er/RosaDHMG08}
Marcello {La Rosa}, Marlon Dumas, Arthur H.~M. ter Hofstede, Jan Mendling, and
  Florian Gottschalk.
\newblock Beyond control-flow: Extending business process configuration to
  roles and objects.
\newblock In \emph{Conceptual Modeling - {ER} 2008, 27th International
  Conference on Conceptual Modeling, Barcelona, Spain, October 20-24, 2008.
  Proceedings}, volume 5231 of \emph{Lecture Notes in Computer Science}, pages
  199--215. Springer, 2008.
\newblock ISBN 978-3-540-87876-6.

\bibitem[List and Korherr(2005)]{DBLP:conf/er/ListK05}
Beate List and Birgit Korherr.
\newblock A {UML} 2 profile for business process modelling.
\newblock In Jacky Akoka, Stephen~W. Liddle, Il{-}Yeol Song, Michela
  Bertolotto, Isabelle Comyn{-}Wattiau, Samira~Si{-}Said Cherfi, Willem{-}Jan
  van~den Heuvel, Bernhard Thalheim, Manuel Kolp, Paolo Bresciani, Juan
  Trujillo, Christian Kop, and Heinrich~C. Mayr, editors, \emph{Perspectives in
  Conceptual Modeling, {ER} 2005 Workshops AOIS, BP-UML, CoMoGIS, eCOMO, and
  QoIS, Klagenfurt, Austria, October 24-28, 2005, Proceedings}, volume 3770 of
  \emph{Lecture Notes in Computer Science}, pages 85--96. Springer, 2005.

\bibitem[List and Korherr(2006)]{DBLP:conf/sac/ListK06}
Beate List and Birgit Korherr.
\newblock An evaluation of conceptual business process modelling languages.
\newblock In \emph{Proceedings of the 2006 {ACM} Symposium on Applied Computing
  (SAC), Dijon, France, April 23-27, 2006}, pages 1532--1539. {ACM}, 2006.

\bibitem[Loniewski et~al.(2010)Loniewski, Insfran, and
  Abrah{\~a}o]{Loniewski2010ASR}
Grzegorz Loniewski, Emilio Insfran, and Silvia~Mara Abrah{\~a}o.
\newblock A systematic review of the use of requirements engineering techniques
  in model-driven development.
\newblock In \emph{Proceedings of the International Conference on Model Driven
  Engineering Languages and Systems (MODELS 2010)}, volume 6395 of
  \emph{Lecture Notes in Computer Science}, pages 213--227. Springer, 2010.

\bibitem[{Mahdi} et~al.(2012){Mahdi}, {Elmansouri}, and {Chaoui}]{Mahdi2012}
K.~{Mahdi}, R.~{Elmansouri}, and A.~{Chaoui}.
\newblock Transforming business patterns to colored petri nets using graph
  grammars.
\newblock In \emph{2012 22nd International Conference on Computer Theory and
  Applications (ICCTA)}, pages 72--78, 2012.

\bibitem[Martins and Zacarias(2015)]{martins2015business}
Paula~Ventura Martins and Marielba Zacarias.
\newblock Business process and practice alignment meta-model.
\newblock \emph{Procedia Computer Science}, 64:\penalty0 314--323, 2015.

\bibitem[Mendling(2008)]{Mendling2008}
Jan Mendling.
\newblock Event-driven process chains (epc).
\newblock In \emph{{Metrics for Process Models}}, volume~6 of \emph{Lecture
  Notes in Business Information Processing}, pages 17--57. Springer Berlin
  Heidelberg, 2008.

\bibitem[Mertens et~al.(2017)Mertens, Gailly, and
  Poels]{DBLP:journals/eswa/MertensGP17}
Steven Mertens, Frederik Gailly, and Geert Poels.
\newblock Towards a decision-aware declarative process modeling language for
  knowledge-intensive processes.
\newblock \emph{Expert Syst. Appl.}, 87:\penalty0 316--334, 2017.

\bibitem[Meyer and Weske(2013)]{DBLP:conf/bpm/0001W13}
Andreas Meyer and Mathias Weske.
\newblock Activity-centric and artifact-centric process model roundtrip.
\newblock In \emph{Business Process Management Workshops - {BPM} 2013
  International Workshops, Beijing, China, August 26, 2013, Revised Papers},
  volume 171 of \emph{Lecture Notes in Business Information Processing}, pages
  167--181. Springer, 2013.
\newblock ISBN 978-3-319-06256-3.
\newblock \doi{10.1007/978-3-319-06257-0}.

\bibitem[Milanovi{\'{c}} et~al.(2008)Milanovi{\'{c}}, Ga{\v{s}}evi{\'{c}}, and
  Wagner]{DBLP:conf/edoc/MilanovicGW08}
Milan Milanovi{\'{c}}, Dragan Ga{\v{s}}evi{\'{c}}, and Gerd Wagner.
\newblock Combining rules and activities for modeling service-based business
  processes.
\newblock In Marten van Sinderen, Jo{\~{a}}o Paulo~A. Almeida,
  Lu{\'{\i}}s~Ferreira Pires, and Maarten Steen, editors, \emph{Workshops
  Proceedings of the 12th International {IEEE} Enterprise Distributed Object
  Computing Conference, {ECOCW} 2008, 16 September 2008, Munich, Germany},
  pages 11--22. {IEEE} Computer Society, 2008.

\bibitem[Mili et~al.(2010)Mili, Tremblay, Jaoude, Lefebvre, Elabed, and
  El{-}Boussaidi]{DBLP:journals/csur/MiliTJLEE10}
Hafedh Mili, Guy Tremblay, Guitta~Bou Jaoude, {\'{E}}ric Lefebvre, Lamia
  Elabed, and Ghizlane El{-}Boussaidi.
\newblock Business process modeling languages: Sorting through the alphabet
  soup.
\newblock \emph{{ACM} Comput. Surv.}, 43\penalty0 (1):\penalty0 4:1--4:56,
  2010.

\bibitem[Momotko and Subieta(2004)]{DBLP:conf/adbis/MomotkoS04}
Mariusz Momotko and Kazimierz Subieta.
\newblock Process query language: {A} way to make workflow processes more
  flexible.
\newblock In \emph{Advances in Databases and Information Systems, 8th East
  European Conference, {ADBIS} 2004, Budapest, Hungary, September 22-25, 2004,
  Proceedings}, volume 3255 of \emph{Lecture Notes in Computer Science}, pages
  306--321. Springer, 2004.
\newblock ISBN 3-540-23243-5.

\bibitem[M{\"u}ller et~al.(2007)M{\"u}ller, Reichert, and
  Herbst]{10.1007/978-3-540-76848-7_10}
Dominic M{\"u}ller, Manfred Reichert, and Joachim Herbst.
\newblock Data-driven modeling and coordination of large process structures.
\newblock In Robert Meersman and Zahir Tari, editors, \emph{On the Move to
  Meaningful Internet Systems 2007: CoopIS, DOA, ODBASE, GADA, and IS}, volume
  4803 of \emph{Lecture Notes in Computer Science}, pages 131--149, Berlin,
  Heidelberg, 2007. Springer Berlin Heidelberg.
\newblock ISBN 978-3-540-76848-7.

\bibitem[{Murata}(1989)]{Murata89}
T.~{Murata}.
\newblock Petri nets: Properties, analysis and applications.
\newblock \emph{Proceedings of the IEEE}, 77\penalty0 (4):\penalty0 541--580,
  1989.

\bibitem[Natschl{\"{a}}ger(2011)]{DBLP:conf/bpmn/Natschlager11}
Christine Natschl{\"{a}}ger.
\newblock Towards a {BPMN} 2.0 ontology.
\newblock In \emph{Business Process Model and Notation - Third International
  Workshop, {BPMN} 2011, Lucerne, Switzerland, November 21-22, 2011.
  Proceedings}, volume~95 of \emph{Lecture Notes in Business Information
  Processing}, pages 1--15. Springer, 2011.
\newblock ISBN 978-3-642-25159-7.

\bibitem[Nelson et~al.(2012)Nelson, Poels, Genero, and Piattini]{Nelson2012}
H.~James Nelson, Geert Poels, Marcela Genero, and Mario Piattini.
\newblock A conceptual modeling quality framework.
\newblock \emph{Software Quality Journal}, 20\penalty0 (1):\penalty0 201--228,
  2012.
\newblock \doi{10.1007/s11219-011-9136-9}.

\bibitem[Nguyen et~al.(2015)Nguyen, Kramer, Klein, and
  Traon]{Nguyen:2015:ESR:2831506.2831600}
Phu~H. Nguyen, Max Kramer, Jacques Klein, and Yves~Le Traon.
\newblock An extensive systematic review on the model-driven development of
  secure systems.
\newblock \emph{Inf. Softw. Technol.}, 68\penalty0 (C):\penalty0 62--81,
  December 2015.
\newblock ISSN 0950-5849.

\bibitem[{Object Management Group}(2011)]{model2011notation}
{Object Management Group}.
\newblock {Business Process Model and Notation (BPMN) version 2.0}.
\newblock https://www.omg.org/spec/BPMN/, 2011.

\bibitem[Ouali et~al.(2016)Ouali, Mhiri, and
  Bouzguenda]{DBLP:conf/kes/OualiMB16}
Sonya Ouali, Mohamed Mhiri, and Lotfi Bouzguenda.
\newblock A multidimensional knowledge model for business process modeling.
\newblock In Robert~J. Howlett, Lakhmi~C. Jain, Bogdan Gabrys, Carlos Toro, and
  Chee~Peng Lim, editors, \emph{Knowledge-Based and Intelligent Information
  {\&} Engineering Systems: Proceedings of the 20th International Conference
  KES-2016, York, UK, 5-7 September 2016}, volume~96 of \emph{Procedia Computer
  Science}, pages 654--663. Elsevier, 2016.

\bibitem[Papavassiliou and Mentzas(2003)]{DBLP:journals/jkm/PapavassiliouM03}
Giorgos Papavassiliou and Gregoris Mentzas.
\newblock Knowledge modelling in weakly-structured business processes.
\newblock \emph{J. Knowledge Management}, 7\penalty0 (2):\penalty0 18--33,
  2003.

\bibitem[Pesic et~al.(2007)Pesic, Schonenberg, and van~der
  Aalst]{Pesic:2007:DFS:1317532.1318056}
Maja Pesic, Helen Schonenberg, and Wil M.~P. van~der Aalst.
\newblock Declare: Full support for loosely-structured processes.
\newblock In \emph{Proceedings of the 11th IEEE International Enterprise
  Distributed Object Computing Conference}, EDOC '07, pages 287--287,
  Washington, DC, USA, 2007. IEEE Computer Society.
\newblock ISBN 0-7695-2891-0.

\bibitem[Ramdoyal et~al.(2013)Ramdoyal, Ponsard, Derbali, Schwanen, Linden, and
  Jacquet]{DBLP:conf/iceis/RamdoyalPDSLJ13}
Ravi Ramdoyal, Christophe Ponsard, Myriam{-}Amina Derbali, Gabriel Schwanen,
  Isabelle Linden, and Jean{-}Marie Jacquet.
\newblock A generic workflow metamodel to support resource-aware decision
  making.
\newblock In \emph{Proceedings of the 15th International Conference on
  Enterprise Information Systems - Volume 1: ICEIS}, pages 243--250.
  SciTePress, 2013.

\bibitem[Redding et~al.(2009)Redding, Dumas, ter Hofstede, and
  Iordachescu]{DBLP:conf/wecwis/ReddingDHI09}
Guy Redding, Marlon Dumas, Arthur H.~M. ter Hofstede, and Adrian Iordachescu.
\newblock Modelling flexible processes with business objects.
\newblock In Birgit Hofreiter and Hannes Werthner, editors, \emph{2009 {IEEE}
  Conference on Commerce and Enterprise Computing, {CEC} 2009, Vienna, Austria,
  July 20-23, 2009}, pages 41--48. {IEEE} Computer Society, 2009.

\bibitem[Rittgen(2006)]{DBLP:journals/ejis/Rittgen06}
Peter Rittgen.
\newblock A language-mapping approach to action-oriented development of
  information systems.
\newblock \emph{Eur. J. Inf. Syst.}, 15\penalty0 (1):\penalty0 70--81, 2006.

\bibitem[Rosemann et~al.(2008)Rosemann, Recker, and
  Flender]{DBLP:journals/ijbpim/RosemannRF08}
Michael Rosemann, Jan Recker, and Christian Flender.
\newblock Contextualisation of business processes.
\newblock \emph{{International Journal of Business Process Integration and
  Management}}, 3\penalty0 (1):\penalty0 47--60, 2008.

\bibitem[Ruiz et~al.(2014)Ruiz, Costal, Espa{\~{n}}a, Franch, and
  Pastor]{DBLP:conf/caise/RuizCEFP14}
Marcela Ruiz, Dolors Costal, Sergio Espa{\~{n}}a, Xavier Franch, and
  {\'{O}}scar Pastor.
\newblock Integrating the goal and business process perspectives in information
  system analysis.
\newblock In \emph{Advanced Information Systems Engineering - 26th
  International Conference, CAiSE 2014, Thessaloniki, Greece, June 16-20, 2014.
  Proceedings}, volume 8484 of \emph{Lecture Notes in Computer Science}, pages
  332--346. Springer, 2014.
\newblock ISBN 978-3-319-07880-9.

\bibitem[Russell et~al.(2016)Russell, van~der Aalst, and ter
  Hofstede]{N.-Russell:2016aa}
N.~Russell, W.M.P. van~der Aalst, and A.H.M. ter Hofstede.
\newblock \emph{Workflow Patterns: The Definitive Guide}.
\newblock MIT Press, 2016.

\bibitem[Russell et~al.(2005)Russell, van~der Aalst, ter Hofstede, and
  Edmond]{DBLP:conf/caise/RussellAHE05}
Nick Russell, Wil M.~P. van~der Aalst, Arthur H.~M. ter Hofstede, and David
  Edmond.
\newblock Workflow resource patterns: Identification, representation and tool
  support.
\newblock In \emph{Advanced Information Systems Engineering, 17th International
  Conference, CAiSE 2005, Porto, Portugal, June 13-17, 2005, Proceedings},
  volume 3520 of \emph{Lecture Notes in Computer Science}, pages 216--232.
  Springer, 2005.
\newblock ISBN 3-540-26095-1.

\bibitem[Santiago et~al.(2012)Santiago, Jim{\'e}nez, Vara, {De Castro},
  Bollati, and Marcos]{Santiago2012ModelDrivenEA}
Iv{\'a}n Santiago, {\'A}lvaro Jim{\'e}nez, Juan~M. Vara, Valeria {De Castro},
  Ver{\'o}nica~Andrea Bollati, and Esperanza Marcos.
\newblock Model-driven engineering as a new landscape for traceability
  management: A systematic literature review.
\newblock \emph{Information {\&} Software Technology}, 54:\penalty0 1340--1356,
  2012.

\bibitem[{Santos Jr.} et~al.(2010){Santos Jr.}, Almeida, and
  Guizzardi]{DBLP:conf/sac/SantosAG10}
Paulo~S{\'{e}}rgio {Santos Jr.}, Jo{\~{a}}o Paulo~A. Almeida, and Giancarlo
  Guizzardi.
\newblock An ontology-based semantic foundation for {ARIS} {EPCs}.
\newblock In \emph{Proceedings of the 2010 {ACM} Symposium on Applied Computing
  (SAC), Sierre, Switzerland, March 22-26, 2010}, pages 124--130. {ACM}, 2010.
\newblock ISBN 978-1-60558-639-7.

\bibitem[{Santos Jr.} et~al.(2011){Santos Jr.}, Almeida, and
  Lavarezi~Pianissolla]{IJBPIM-ARIS}
Paulo~S{\'{e}}rgio {Santos Jr.}, Jo{\~a}o Almeida, and Thiago
  Lavarezi~Pianissolla.
\newblock Uncovering the organisational modelling and business process
  modelling languages in the aris method.
\newblock \emph{Int. J. of Business Process Integration and Management},
  5:\penalty0 130 -- 143, 05 2011.

\bibitem[Scheer(2013)]{scheer2013aris}
August-Wilhelm Scheer.
\newblock \emph{ARIS---vom Gesch{\"a}ftsprozess zum Anwendungssystem}.
\newblock Springer-Verlag, 2013.

\bibitem[S{\"{o}}derstr{\"{o}}m et~al.(2002)S{\"{o}}derstr{\"{o}}m, Andersson,
  Johannesson, Perjons, and Wangler]{DBLP:conf/caise/SoderstromAJPW02}
Eva S{\"{o}}derstr{\"{o}}m, Birger Andersson, Paul Johannesson, Erik Perjons,
  and Benkt Wangler.
\newblock Towards a framework for comparing process modelling languages.
\newblock In \emph{Advanced Information Systems Engineering, 14th International
  Conference, CAiSE 2002, Toronto, Canada, May 27-31, 2002, Proceedings},
  volume 2348 of \emph{Lecture Notes in Computer Science}, pages 600--611.
  Springer, 2002.
\newblock ISBN 3-540-43738-X.

\bibitem[Sprovieri and Vogler(2015)]{DBLP:conf/es/SprovieriV15}
Danillo Sprovieri and Sandro Vogler.
\newblock Run-time composition of partly structured business processes using
  heuristic planning.
\newblock In \emph{International Conference on Enterprise Systems, {ES} 2015,
  Basel, Switzerland, October 14-15, 2015}, pages 225--232. {IEEE}, 2015.
\newblock \doi{10.1109/ES.2015.30}.

\bibitem[Stratigaki et~al.(2016)Stratigaki, Nikolaidou, Loucopoulos, and
  Anagnostopoulos]{DBLP:conf/wecwis/StratigakiNLA16}
Christina Stratigaki, Mara Nikolaidou, Pericles Loucopoulos, and Dimosthenis
  Anagnostopoulos.
\newblock Business process elicitation from regulatory compliance documents: An
  e-government case study.
\newblock In Elena Kornyshova, Geert Poels, and Christian Huemer, editors,
  \emph{18th {IEEE} Conference on Business Informatics, {CBI} 2016, 29th August
  - 1st September 2016, Paris, France, Volume 2 - Short and Workshop Papers},
  pages 8--13. {IEEE}, 2016.

\bibitem[Strembeck and Mendling(2011)]{DBLP:journals/infsof/StrembeckM11}
Mark Strembeck and Jan Mendling.
\newblock Modeling process-related {RBAC} models with extended {UML} activity
  models.
\newblock \emph{Information {\&} Software Technology}, 53\penalty0
  (5):\penalty0 456--483, 2011.

\bibitem[Stroppi et~al.(2011)Stroppi, Chiotti, and Villarreal]{Stroppi2011}
L.J.R. Stroppi, O.~Chiotti, and P.D. Villarreal.
\newblock A {{BPMN}} 2.0 extension to define the resource perspective of
  business process models.
\newblock \emph{14th Ibero-American Conference on Software Engineering and 14th
  Workshop on Requirements Engineering, CIbSE 2011}, pages 25--38, 2011.

\bibitem[ter Hofstede et~al.(2010)ter Hofstede, van~der Aalst, Adams, and
  Russell]{YAWL:2010aa}
Arthur~H.M. ter Hofstede, Wil~M.P. van~der Aalst, Michael Adams, and Nick
  Russell, editors.
\newblock \emph{{Modern Business Process Automation: YAWL and its Support
  Environment}}.
\newblock Springer, 2010.

\bibitem[Thabet et~al.(2015)Thabet, Ghannouchi, and Ben~Ghezala]{Thabet2015}
D.~Thabet, S.A. Ghannouchi, and H.H. Ben~Ghezala.
\newblock Business process model extension with cost perspective based on
  process mining - cost data description and analysis.
\newblock \emph{Proceedings of the 26th International Business Information
  Management Association Conference - Innovation Management and Sustainable
  Economic Competitive Advantage: From Regional Development to Global Growth,
  IBIMA 2015}, pages 44--58, 2015.

\bibitem[Thom et~al.(2005)Thom, Iochpe, and Mitschang]{thom2005improving}
Lucin{\'{e}}ia~Heloisa Thom, Cirano Iochpe, and Bernhard Mitschang.
\newblock Improving workflow project quality via business process patterns
  based on organizational structure aspects.
\newblock In \emph{Proceedings of the Second GI-Workshop {XML4BPM} - {XML} for
  Business Process Management held at the 11th Conference Business,
  Technologie, and Web {(} {BTW} 2005 {)} Karlsruhe (Germany), 01 March 2005.},
  volume 145 of \emph{{CEUR} Workshop Proceedings}, pages 65--80. CEUR-WS.org,
  2005.

\bibitem[van~der Aalst and Kumar(2001)]{10.1016/S0169-023X(01)00034-9}
W.~M.~P. van~der Aalst and A.~Kumar.
\newblock A reference model for team-enabled workflow management systems.
\newblock \emph{Data Knowl. Eng.}, 38\penalty0 (3):\penalty0 335–363,
  September 2001.
\newblock ISSN 0169-023X.

\bibitem[Wand and Weber(1990)]{Wand:1990ab}
Y.~Wand and R~Weber.
\newblock An ontological model of an information system.
\newblock \emph{IEEE Transactions on Software Engineering}, 16\penalty0
  (11):\penalty0 1282--1292, 1990.

\bibitem[Weigand et~al.(2006)Weigand, Johannesson, Andersson, Bergholtz,
  Edirisuriya, and Ilayperuma]{DBLP:conf/caise/WeigandJABEI06}
Hans Weigand, Paul Johannesson, Birger Andersson, Maria Bergholtz, Ananda
  Edirisuriya, and Tharaka Ilayperuma.
\newblock On the notion of value object.
\newblock In \emph{Advanced Information Systems Engineering, 18th International
  Conference, CAiSE 2006, Luxembourg, Luxembourg, June 5-9, 2006, Proceedings},
  volume 4001 of \emph{Lecture Notes in Computer Science}, pages 321--335.
  Springer, 2006.
\newblock ISBN 3-540-34652-X.

\bibitem[Wei{\ss} and Winkelmann(2011)]{DBLP:conf/hicss/WeissW11a}
Burkhard Wei{\ss} and Axel Winkelmann.
\newblock A metamodel based perspective on the adaptation of a semantic
  business process modeling language to the financial sector.
\newblock In \emph{44th Hawaii International International Conference on
  Systems Science {(HICSS-44} 2011), Proceedings, 4-7 January 2011, Koloa,
  Kauai, HI, {USA}}, pages 1--10. {IEEE} Computer Society, 2011.
\newblock ISBN 978-0-7695-4282-9.

\bibitem[Weske(2012)]{DBLP:books/daglib/0029914}
Mathias Weske.
\newblock \emph{Business Process Management - Concepts, Languages,
  Architectures, 2nd Edition}.
\newblock Springer, 2012.
\newblock ISBN 978-3-642-28615-5.

\bibitem[{Working Group I11}(2006)]{r2ml}
{Working Group I11}.
\newblock Rewerse i1 rule markup language (r2ml)., 2006.
\newblock
  \url{http://www.macs.hw.ac.uk/bisel/rewerse/I1/oxygen.informatik.tu-cottbus.de/rewerse-i1/%40q%3dnode_2f6.html}.

\bibitem[Yahya et~al.(2016)Yahya, Boukadi, Maamar, and
  Ben{-}Abdallah]{DBLP:conf/iiwas/YahyaBMB16}
Fadwa Yahya, Khouloud Boukadi, Zakaria Maamar, and Han{\^{e}}ne Ben{-}Abdallah.
\newblock Towards a meta-modeling approach for social business process
  requirements capture.
\newblock In Gabriele Anderst{-}Kotsis, editor, \emph{Proceedings of the 18th
  International Conference on Information Integration and Web-based
  Applications and Services, iiWAS 2016, Singapore, November 28-30, 2016},
  pages 345--354. {ACM}, 2016.

\bibitem[zur Muehlen(2004)]{DBLP:journals/itm/Muehlen04}
Michael zur Muehlen.
\newblock Organizational management in workflow applications - issues and
  perspectives.
\newblock \emph{Inf. Technol. Manag.}, 5\penalty0 (3-4):\penalty0 271--291,
  2004.
\newblock \doi{10.1023/B:ITEM.0000031582.55219.2b}.

\end{thebibliography}
\end{document}